\documentclass[nofootinbib,amsfonts,prd,aps,12pt]{revtex4}

\usepackage[utf8]{inputenc}
\usepackage[T1]{fontenc}
\usepackage{amsmath,amssymb}
\usepackage{hyperref}
\usepackage{graphicx}
\usepackage{subfigure}

\begin{document}
	
\title{Brief review on black hole loop quantization}
\author{Javier Olmedo}

\affiliation{Department of Physics and Astronomy, Louisiana State University, Baton Rouge, Louisiana 70803-4001, USA}

\begin{abstract}
Here, we present a review about the quantization of spherically-symmetric spacetimes adopting loop quantum gravity techniques. Several models that have been studied so far share similar properties: the resolution of the classical singularity and some of them an intrinsic discretization of the geometry. We also explain the extension to Reissner---Nordstr\"om black holes. Besides, we review how quantum test fields on these quantum geometries allow us to study phenomena, like the Casimir effect or Hawking radiation. Finally, we briefly describe a recent proposal that incorporates spherically-symmetric matter, discussing its relevance for the understanding of black hole evolution.
\end{abstract}

\maketitle
\vfill\break
\tableofcontents
\vfill\break



\section{Introduction}

Black holes are one of the most fascinating entities in the cosmos. Some of their properties can be described with Einstein's theory of gravity in a relatively simple way. However, there are several aspects that are currently not well understood, since it is expected they will require an extension of the classical theory. For instance, one of them is the issue of the classical singularity present in general relativity. Hopefully, a quantum theory of gravity is the natural extension of the classical one there, providing a regular description and free of divergences. Moreover, the evolution of black holes is not completely well understood at present, since it is conceptually and technically involved. Particularly, the last instants of a black hole, where the evaporation process takes place and the Planck regime is expected to be relevant. In this case, besides, the information loss paradox is a fundamental question that is expected to be solved once the problem is understood beyond the semiclassical approximation that has been traditionally adopted. 

Among the different approaches looking for a quantum theory of gravity, loop quantum gravity is one of the most promising ones \cite{lqg,lqg1,lqg2}. It is a nonperturbative, background independent, canonical quantization of general relativity. Its present status is mathematically well defined, but he physical sector is not completely understood, in particular its semiclassical regime. The quantization techniques of loop quantum gravity have been successfully applied in many mini- and midi-superspace models \cite{lqc,lqc1,lqc2}. In those cases where the physical sector has been explicitly solved, this quantization program provides a good semiclassical limit reproducing general relativity and a robust resolution of the classical singularity (replaced by a regular region of high curvature) at the deep Planck regime. In addition, in more general scenarios, where there is no complete understanding of the quantum dynamics, there is strong evidence that several singularities of general relativity are not present \cite{param}. Concerning black hole physics, the study of statistical properties of isolated horizons \cite{iso1,iso2} out of microscopic states is an interesting application of loop quantum gravity. In this case, one can consider an $SU(2)$ Chern–Simons theory in order to construct the microscopic states of the Schwarzschild black hole and compute its entropy. It turns out to be proportional to the area of the horizon, where the proportionality coefficient is a function of the Barbero–Immirzi parameter \cite{lqg,lqg1,lqg2}.

In addition, the quantization of spherically-symmetric spacetimes is receiving special attention at present due to its significance in black hole physics. Einstein's theory of gravity in vacuum gives a suitable description of some of the fundamental aspects of these entities. Indeed, they are connected with fundamental problems in theoretical physics nowadays. For instance, we do not know which is the true physical description close to the classical singularity (where general relativity breaks down) or the essential nature of Hawking radiation. In the latter case, it leads to black hole evaporation and eventually to the information loss paradox. It is commonly believed that all these problems would be successfully solved if they are treated within a quantum theory of gravity. Although loop quantum gravity is not able at present to deal directly with these problems, we believe that it should be able to answer some of these questions within midi-superspace quantizations. For instance, the replacement of the classical singularity by a high curvature region has been noticed by several authors \cite{bojo,ash-bojo,bs,cgp,cgp2,chiou,gp-lett,gop,gop2,cs} within the context of a loop quantization of spherically-symmetric spacetimes.
Several phenomena in quantum field theories on these quantum spacetimes have been studied, as well, like the Casimir effect \cite{gop-casimir} and Hawking radiation \cite{gp-semi}.  It should also be noticed the notable effort at present in order to generalize the previous models closer to the full theory \cite{radial,radial1,radial2}, with special attention to the closely related studies about loop quantum cosmology emerging from group field theories \cite{gpt,gpt1,gpt2}. 

A full quantum description in vacuum following the Dirac quantization approach was carried out in Ref. \cite{gp-lett}. There, the classical constraint algebra is modified in such a way that the Hamiltonian constraint has an Abelian algebra with itself. Then, one adopts the quantization program of loop quantum gravity. It is also possible to find the physical space of solutions and Dirac observables. An additional prescription in vacuum was considered in Ref. \cite{gop}, such that the quantization seems more natural in order to incorporate matter. Several questions related to the dynamics of the vacuum model were considered in Ref. \cite{gop2}. It should be important to notice the challenge of facing the quantization of this kind of models regarding the anomalies present in the constraint algebra, even if one adheres to the Abelianized version of the scalar constraint \cite{bb,bbr}. However, it has been possible to overcome some of these problems in a recent publication \cite{cgop}. The model under consideration is a spherically-symmetric self-gravitating null dust shell. The combination of a standard representation for the shell with a loop quantization for the geometry provides a constraint algebra free of anomalies. 

In this manuscript we will review all those results. We will start with a classical description in Sec. \ref{sec:class-syst}. We will also explain here the Abelianization procedure for the Hamiltonian constraint. In Sec. \ref{sec:kinemat} we will provide the kinematical quantum representation usually adopted in this kind of models. The quantum scalar constraint will be given in Sec. \ref{sec:scalar-const} and its solutions in Sec. \ref{sec:solut}. Finally, in Sec. \ref{sec:physical} one can find a complete physical picture of the model. We will also compare this quantization with other approaches in Sec. \ref{sec:comp}, and its extension for charged black holes will be discussed in Sec. \ref{sec:BH-MQ}. We comment on several phenomena of quantum test fields on these quantum geometries in Sec. \ref{test-f}, like the Casimir and the Hawking effects. In addition, the quantization in presence of a thin shell will be summarized in Sec. \ref{sec:shell}. Sec. \ref{sec:concl} is dedicated to the conclusions.

\section{Classical system}\label{sec:class-syst}

\subsection{Kinematics}

Classical gravity in Ashtekar---Barbero variables with spherical symmetry has been studied in Refs. \cite{bojo,bs} (a more general discussion can be found in Refs. \cite{bencorteit,bengtsson,bojo2}). In this reduced theory, we are left with a $1+1$ model described by three pairs of canonical variables and three first class constraints. One of them is a residual Gauss constraint generating $U(1)$ transformations. Its implementation at the quantum level is well known \cite{bs,cgp,chiou}. For the sake of simplicity, we will adopt a description in terms of gauge-invariant quantities and we will carry out a classical gauge fixing in order to eliminate the Gauss constraint. The usual strategy is to identify the corresponding gauge-invariant quantities after a series of canonical transformations that separate the pure gauge degrees of freedom from those fields that commute with the Gauss constraint (gauge invariant fields). At the end, we are left with the gauge invariant variables
\begin{align}\nonumber\label{eq:poiss}
&\{K_x(x),E^x(\tilde x)\}=G\delta(x-\tilde x),\\
&\{{K}_\varphi(x),E^\varphi(\tilde x)\}=G\delta(x-\tilde x),
\end{align} 
where $G$ is the Newton constant. The quantities $E^x$ and $E^\varphi$ have a clear geometrical meaning, since they determine the components of the spatial metric 
\begin{equation}
ds^2=\frac{(E^\varphi)^2}{|E^x|}dx^2+|E^x| (d\theta^2+\sin^2\theta d\phi^2).
\end{equation}
On the other hand, $K_x$ and $K_\varphi$ represent, in triadic form, the components the extrinsic curvature. 

The other two constraints correspond to the diffeomorphism and the scalar constraints
\begin{subequations}
	\begin{align}
	& H_r:=G^{-1}[E^\varphi K_\varphi'-(E^x)' K_x]\,,\label{eq:difeo}\\ \nonumber
	&H :=G^{-1}\left\{\frac{\left[(E^x)'\right]^2}{8\sqrt{E^x}E^\varphi}
	-\frac{E^\varphi}{2\sqrt{E^x}} - 2 K_\varphi \sqrt{E^x} K_x  
	-\frac{E^\varphi K_\varphi^2}{2 \sqrt{E^x}}\right.\\
	&\left.-\frac{\sqrt{E^x}(E^x)' (E^\varphi)'}{2 (E^\varphi)^2} +
	\frac{\sqrt{E^x} (E^x)''}{2 E^\varphi}\right\}\,,\label{eq:scalar1}
	\end{align}
\end{subequations}
respectively. Here, the primes mean derivation with respect to $x$. One can easily check that the constraint algebra under Poisson brackets is
\begin{subequations}
	\begin{align}
	&\{H_r(N_r),H_r(\tilde N_r)\}=H_r(N_r\tilde N_r'-N_r'\tilde N_r),\\
	&\{H(N),H_r(N_r)\}=H(N_r N'),\\
	&\{H(N),H(\tilde N)\}=H_r\left(\frac{E^x}{(E^\varphi)^2}\left[N\tilde N'-N' \tilde N\right]\right).
	\end{align}
\end{subequations}
We observe that it is equipped with structure functions (like the full theory), with the
ensuing difficulties for achieving a consistent quantization \cite{haji-kuch,komar1,komar2,brahma}. 

The dynamics of the system is ruled by the Hamiltonian 
\begin{equation}\label{eq:total-ham}
H_T=\int dx (NH+N_rH_r).
\end{equation}

The equations of motion can be computed and the traditional solutions to Einstein's equations found, assuming suitable falloff conditions (they will be explained below).

\subsection{Weak Dirac observables}

This classical theory possesses one weak Dirac observable: the Arnowitt---Deser---Misner (ADM) mass of the black hole. Its Poisson bracket with the diffeomorphism and the scalar constraint vanishes on-shell. One can easily realize that there will be no other Dirac observables. Although a detailed analysis can be carried out, we will explain here heuristically the necessary arguments to reach this conclusion. There are two local degrees of freedom (four phase space variables) and two local first class constraints. If we divide the phase space  into the homogeneous and inhomogeneous sectors, the latter will be constrained by two local first class constraints. No physical degrees of freedom are left. However, the homogeneous sector, with four homogeneous phase space functions, is constrained only by the homogeneous Hamiltonian constraint. In this sector, the diffeomorphism constraint vanishes identically. Then, we conclude that there will be left one and only one unconstrained global degree of freedom in this vacuum theory.

Let us define the following phase space functions
\begin{align}
&{\cal M}(x):=\frac{2E^x\sqrt{E^x} K_x K_\varphi}{GE^\varphi}+\frac{\sqrt{E^x}E^x(E^x)' (E^\varphi)'}{2G (E^\varphi)^3} -
\frac{\sqrt{E^x} E^x(E^x)''}{2 G(E^\varphi)^2},\\
&\tilde {\cal M}(x):=-\frac{1}{2G}\sqrt{E^x}(1+K_\varphi^2) +\frac{\sqrt{E^x}[(E^x)']^2}{8G(E^\varphi)^{2}}.
\end{align}
One can see, after some calculations, that $\dot{{\cal M}}(x)= 0$ on-shell (where the dot is derivation with respect to time, and it is obtained after taking Poisson brackets with the total Hamiltonian), concluding that ${\cal M}(x)$ is a weak Dirac observable. It corresponds to the ADM mass ${{\cal M}}(x)= M$ of the system. Concerning the phase space function $\tilde {\cal M}(x)$, it is possible to see that $\tilde {\cal M}(x)={\cal M}(x)-E^x H/E^\varphi$. Therefore, it is a linear combination of the weak observable ${\cal M}(x)$ and the scalar constraint $H$. We then conclude that ${\dot{\tilde{\cal M}}}(x)= 0$ on the constraint surface. In addition, $\tilde {\cal M}(x)={\cal M}(x)$ on-shell, so there is only one linearly-independent Dirac observable, and consequently, only one physical degree of freedom, in agreement with the previous arguments.

\subsection{Abelianization of the Hamiltonian constraint}

A quantum description of this model within the Dirac quantization approach requires dealing with a consistent quantum constraint algebra free of anomalies. The presence of structure functions makes this task even more difficult. Indeed, it is well known that one has to deal with non-self-adjoint constraints \cite{haji-kuch,komar1,komar2,brahma}. Besides, if one follows a quantum representation within loop quantum gravity, it is necessary to face 
additional anomalies with respect to a standard representation (see, for instance, Refs. \cite{bb,bbr,brahma,bojo3}), like the emergence of discrete structures not present in the continuum theory. 

In recent proposals, a redefinition of the scalar constraint has been adopted at the classical level that provides an Abelian Poisson algebra of the new scalar constraint with itself. In totally constrained theories, the original set of first class constraints can be transformed into a new one such that the new constraint algebra is Abelian \cite{hennteit}. Nevertheless, this strategy is not commonly employed for two reasons: i) the new set of constraints usually has a complicated functional form with respect to the phase space variables in comparison with the original set of constraints, and ii) the Abelianization is only valid locally, in general, i.e. it does not cover the whole constraint surface \cite{hennteit}.

In loop quantum gravity, the implementation of the spatial diffeomorphisms at the quantum level is well known, while the symmetries generated by the scalar constraint are not well understood. For this reason, in this manuscript, we will only adopt a redefinition of the scalar constraint, leaving the diffeomorphism one unaltered. In this case, we can complete the quantization of the vacuum theory, as we will see latter. 

Let us then consider the following linear combination of constraints 
\begin{equation}\label{eq:hnew}
H_{\rm new} :=\frac{\left(E^x\right)'}{E^\varphi}H-2\frac{\sqrt{E^x}}{E^\varphi}K_\varphi H_r=\frac{1}{G}\left[ \sqrt{E^x}\left(1-\frac{[(E^x)']^2 }{4 (E^\varphi)^2}+K_\varphi^2\right)\right]',
\end{equation}
where we call $H_{\rm new}$ the new Abelian scalar constraint. This is tantamount to redefining in the action the lapse and shift functions as \cite{gp-lett}
\begin{equation}
N^{\rm new}_r:= N_r -2 N\frac{K_\varphi\sqrt{E^x}}{\left(E^x\right)'},\quad N_{\rm new} := N \frac{E^\varphi}{\left(E^x\right)'}.
\end{equation}
In these two relations, the homogeneous limit seems to be problematic. This is the typical situation for the study of the interior of the black hole as a Kantowski---Sachs spacetime, either by further symmetry reduction \cite{ash-bojo} or as a gauge fixing condition \cite{cgp2}. However, in this case, one can work with the original set of constraints in Eqs. \eqref{eq:difeo} and \eqref{eq:scalar1}, where the quantization prescriptions adopted so far are free of anomalies, since there, the diffeomorphism constraint vanishes and the scalar one is homogeneous. 

In the following, it will be more convenient to work with an integrated version of $H_{\rm new}$. It requires indeed integrating by parts the new Hamiltonian constraint times the new lapse. However, we need to take into account the fall-off conditions of the fields in order to carry out the integration consistently. 
They have been extensively studied, for instance, in Ref. \cite{kuchar}. For a given initial Cauchy surface with $t=t_0$, the fall-off conditions at both spatial infinities are given by
\begin{align}\nonumber
&E^\varphi(x)=|x|+2GM_{\pm}+{\cal O}(|x|^{-\varepsilon}),\\\nonumber
&E^x(x)=|x|^2+2|x|^{1-\varepsilon}+{\cal O}(|x|^{-2\varepsilon}),\\\nonumber
&K_\varphi(x)={\cal O}(|x|^{-(1+\varepsilon)}),\\
&K_x(x)={\cal O}(|x|^{-(2+\varepsilon)}),
\end{align}
for the phase space variables, while the original lapse and shift functions behave as
\begin{align}\nonumber
&N(x)=N_\pm+{\cal O}(|x|^{-\varepsilon}),\\
&N_r(x)={\cal O}(|x|^{-\varepsilon}).
\end{align}

Here, $M_\pm$ and $N_\pm$ are some functions that only depend on time $t$. Since the new lapse function $N_{\rm new}$ is known in terms of the phase space functions and the original lapse function $N$, the fall-off conditions of $N_{\rm new}$ can be easily obtained. If we smear the constraint in Eq. \eqref{eq:hnew} with the lapse $N_{\rm new}$, and integrating by parts together with the fall-off conditions, we obtain
\begin{equation}
H_{\rm new} (N_{\rm new})=2N_+ M_+-2N_- M_--\frac{1}{G}\int dx N'_{\rm new}\left[ \sqrt{E^x}\left(1-\frac{[(E^x)']^2 }{4 (E^\varphi)^2}+K_\varphi^2\right)\right].
\end{equation}
We may notice that $M_\pm=M$ since both functions $M_\pm$ coincide with the function ${\cal M}(x)$ on-shell (particularly, at $x\to\pm\infty$), and that $(N_+-N_-)=\int dx N_{\rm new}'$. With this in mind, the previous expression can be simplified
\begin{equation}
H_{\rm new} (N_{\rm new})=-\frac{1}{G}\int dx N'_{\rm new}\left[ \sqrt{E^x}\left(1-\frac{[(E^x)']^2 }{4 (E^\varphi)^2}+K_\varphi^2\right)-2GM\right].
\end{equation}
Let us consider an additional redefinition of the scalar constraint by introducing the lapse $\tilde N:=-N_{\rm new}'\sqrt{E^x}(E^\varphi)^{-1}$. Then, we get the following smeared scalar constraint
\begin{equation}\label{eq:H_new-den}
\tilde H(\tilde N) :=\frac{1}{G}\int dx \tilde N
E^\varphi\bigg[ K_\varphi^2-\frac{[(E^x)']^2}{4
	(E^\varphi)^2}+\left(1-\frac{2 G M}{\sqrt{E^x}}\right)\bigg].
\end{equation}

As it was noticed in Ref. \cite{kuchar}, in order to have a well-posed variational problem, the function $M$ must be considered as an additional degree of freedom. It requires the introduction of a boundary term. Otherwise, we would get the undesired equation $\int dx\tilde N E^\varphi/\sqrt{E^x}=(N_+-N_-)=0$, since there is no a priory reason for the coincidence of the lapse functions at the two spatial infinities. The boundary term will be $\int dt M\dot \tau$, such that $\tau$ represents the canonically-conjugated variable of $M$ (the proper time of an observer in the spatial infinity). The final action would be
\begin{equation}
S=\int dt \left[M\dot \tau+\int dx G^{-1}\left(E^\varphi\dot K_\varphi+E^x\dot K_x\right)-\tilde H(\tilde N)-H_r(N_r)\right].
\end{equation}

If we take variations of the action with respect to $\tau$, it provides the condition $\dot M=0$. Variations with respect to $M$ yield $\dot \tau = -\int dx\tilde N E^\varphi/\sqrt{E^x}=-(N_+-N_-)$. This involves the possibility of choosing different time parametrizations at the two spatial infinities. Finally, variations with respect to the remaining phase space functions, lapse and shift yields a set of equations equivalent to the usual Einstein's ones.

\section{Kinematical Hilbert space}\label{sec:kinemat}

Let us proceed with the quantization of the classical theory. Since the mass of the black hole corresponds to a global degree of freedom, we will adopt a quantization for it within a standard description. Among the different choices, we will consider a measure $dM$ compatible with dilations of the mass instead of translations (see, for instance, Ref. \cite{thiem-kast}). In this situation emerges a new observable $\epsilon=\pm1$ corresponding to the sign of the classical mass. We will restrict the study to the sector $\epsilon=1$ where $M>0$, ensuring the positivity of the mass $M$. The representation we adopt here will be ${\cal H}_{\rm kin}^m=L^2(\mathbb{R}^+,M^{-1}dM)$. For the geometry, we adhere to a loop quantum gravity representation \cite{bs}. Here, the basic bricks are one-dimensional graphs determined by the matrix elements of holonomies of $su(2)$-connections along non-overlapping edges $e_j$ connected by vertices $v_j$. In spherically-symmetric spacetimes it is natural to identify the variable $K_x$ with edges of the graph
along the radial direction and $K_\varphi$ with vertices on it. Besides, two edges are joined by means of a vertex. For a given graph $g$, there is a basis of states 
\begin{align*}
|\vec{\mu},\vec{k}\rangle=\mbox{
	\begin{picture}(200,15)(0,0)
	\put(0,5){\line(1,0){200}}
	\put(50,5){\circle*{5}}
	\put(100,5){\circle*{5}}
	\put(150,5){\circle*{5}}
	\put(50,-3){\makebox(0,0){$\mu_{j-1}$}}
	\put(100,-3){\makebox(0,0){$\mu_j$}}
	\put(150,-3){\makebox(0,0){$\mu_{j+1}$}}
	\put(25,10){\makebox(0,0){$\cdots$}}
	\put(75,12){\makebox(0,0){$k_j$}}
	\put(125,12){\makebox(0,0){$k_{j+1}$}}
	\put(175,10){\makebox(0,0){$\cdots$}}
	\end{picture}},
\end{align*}
that in the connection representation take the form
\begin{align}\label{eq:kin-graph}
&\Psi_{g,\vec{k},\vec{\mu}}(K_x,K_\varphi) =\prod_{e_j\in g}
\exp\left(i \gamma k_{j} \int_{e_j} dx\,K_x(x)\right)
\prod_{v_j\in g}
\exp\left(i  \gamma \mu_{j} K_\varphi(v_j) \right),
\end{align}
where the label  $k_j\in\mathbb{Z}$ and $\mu_j\in\mathbb{R}$ are the valences (or coloring) associated with the edge $e_j$ and the vertex $v_j$, respectively. Besides, $\gamma$ is the Immirzi parameter \cite{lqg}.

Let us notice that, due to the symmetry reduction, the component $K_\varphi$ is not a connection, but a scalar. They are usually represented as point holonomies. Therefore, the kinematical representation that mimics the full theory is given in terms of quasiperiodic functions of these connection-like variables, i.e.
$L^2(\mathbb{R}_{\rm Bohr},d\mu_{\rm Bohr})$, with
$\mathbb{R}_{\rm Bohr}$ the Bohr compactification of the real line and $d\mu_{\rm
	Bohr}$ the natural invariant measure under translations on that set. 

The kinematical Hilbert space ${\cal H}_{\rm kin}$ turns out to be the tensor product
\begin{equation}
{\cal H}_{\rm kin}={\cal H}_{\rm kin}^M\otimes\left[\bigotimes_{j}^n\ell^2_j\otimes L_j^2(\mathbb{R}_{\rm Bohr},d\mu_{\rm Bohr})\right],
\end{equation}
which is endowed with the inner product
\begin{align}\label{eq:kin-inner-prod-qp}
\langle\vec{k},\vec{\mu},M
|\vec{k}',\vec{\mu}',M'
\rangle=\delta_{\vec{k},\vec{k}'}\delta_{\vec{\mu},\vec{\mu}'}\delta(M-M')\;.
\end{align}
Here, $\ell^2_j$ means the Hilbert space of square summable functions corresponding to the holonomies of the connection $K_x$ in the radial direction and $n$ the total number of vertices. In addition, we must keep in mind that two states corresponding to different graphs, for instance $g$ and $g'$, are mutually orthogonal.

On this basis, the basic operators acting by multiplication are represented as 
\begin{subequations}
	\begin{align}
	&{\hat{M} } |\vec{k},\vec{\mu},M\rangle
	= M |\vec{k},\vec{\mu},M\rangle,\label{eq:M}\\
	&{\hat{E}^x(x) } |\vec{k},\vec{\mu},M\rangle
	= \gamma \ell_{\rm Pl}^2 k_j |\vec{k},\vec{\mu},M\rangle,\label{eq:ex}
	\\
	&\hat{E}^\varphi(x) |\vec{k},\vec{\mu},M\rangle
	= \gamma \ell_{\rm Pl}^2 \sum_{j}^n \delta\big(x-x_j\big)\mu_j 
	|\vec{k},\vec{\mu},M\rangle,\label{eq:evarphi}
	\end{align}
\end{subequations}
where $k_j$ is the valence of the edge either if $x\in e_j$ or to the vertex to the left of $e_j$. In addition, $x_j$ is the position of the vertex $v_j$.

Since the only component of the connection that is present in the scalar constraint is $K_\varphi(x)$, it will be represented in terms of point holonomies of length $\rho(x)$, i.e.,
\begin{equation}
\hat N^\varphi_{\pm\rho}(x) = \widehat{e^{\pm i  \gamma \rho(x) K_\varphi(x)}}.
\end{equation}
If we write the kinematical states of the basis as
\begin{equation}\label{eq:ket-dec}
|\vec{k},\vec{\mu},M\rangle=\left(\bigotimes_{j}^n|k_j\rangle\otimes|\mu_j\rangle\right)\otimes|M\rangle,
\end{equation}
the operator $\hat N^\varphi_{\rho}(x)$ evaluated at $x=x_j$, i.e., on a given vertex, has the following action
\begin{equation}
\hat N^\varphi_{\pm \rho_j}(x_j) |\mu_j\rangle
= |\mu_j\pm\rho_j\rangle,
\end{equation}
where $\rho_j=\rho(x_j)$. If $x$ is not on a vertex, for instance $x_{j}<x<x_{j+1}$, the action is
\begin{equation}\label{eq:ket-dec2}
\hat N^\varphi_{\pm\rho}(x)|\vec{k},\vec{\mu},M\rangle=\left(\bigotimes_{i<j}^j|k_i\rangle\otimes|\mu_i\rangle\right)\otimes\left(|k_j\rangle\otimes|\pm\rho\rangle\right)\otimes\left(\bigotimes_{i={j+1}}^n|k_i\rangle\otimes|\mu_i\rangle\right)\otimes|M\rangle.
\end{equation}
The new graph has a total number of vertices equal to $n+1$. Evidently, the operator $N^\varphi_{\rho}(x)$ creates a new vertex at $x$ between $x_{j}$ and $x_{j+1}$.

\section{Representation of the Hamiltonian Constraint}\label{sec:scalar-const}

The quantum dynamics of the system is codified in the quantum Hamiltonian constraint. It will be given here by
\begin{align}\label{eq:quant-scalar-constr}
&  \hat{H}(N)=\frac{1}{G}\int dx N(x)\hat P\left\{  
\hat\Theta
-\frac{1}{4}\widehat{\left[\frac{\left[({E}^x)'\right]^2}{{E}^\varphi}\right]} +\hat{E}^\varphi\left(1
-\frac{2 G \hat M}{\sqrt{\hat{E}^x}}\right) \right\}\hat P.
\end{align}
This operator is similar to the one adopted in Ref. \cite{gop}, up to the operator $\hat P$, defined as
\begin{equation}
\hat P|g,\vec{k},\vec{\mu},M\rangle=\prod_{j}^n{\rm sgn}(k_j){\rm sgn}(\mu_j)|g,\vec{k},\vec{\mu},M\rangle,
\end{equation}
where ${\rm sgn}(x)$ is the standard sign function with ${\rm sgn}(0)=0$. 

The quantum constraint has a well-defined action on the kinematical Hilbert space. Besides, it annihilates the normalizable states with any $k_j=0$ and/or $\mu_j=0$. In this situation, we can restrict the study to the orthogonal complement of those states, i.e., all those spin networks with nonvanishing $k_j$ and $\mu_j$. On this subspace, the triads cannot vanish, which has important consequences for the singularity resolution\footnote{Similar arguments were already used in Refs. \cite{bianchi,mmo} for the singularity resolution in cosmological scenarios.}. For instance, the areas of the spheres of symmetry, codified in the spectrum of the operator $\hat E^x(x)$, can never vanish on this subspace. Therefore, the typical spatial singularity at the center of the black hole is naturally regularized (as well as some of the coordinate singularities). Let us mention that the Dirac quantization approach would regard these states annihilated by the constraint as physical states. Nevertheless, they are indeed trivial solutions associated with the trivial representations of the kinematical operators. It is important to keep in mind that they will not contribute to the nontrivial solutions of the constraint. We will then ignore those trivial solutions and restrict the study to their orthogonal complement.

The action of the operator $\hat\Theta(x)$ on the kinematical states is 
\begin{align}
&\hat\Theta(x)|\vec{k},\vec{\mu},M\rangle
= \sum_{j}^n \delta(x-x_j) \hat\Omega_\varphi^2 (x_j)
|\vec{k},\vec{\mu},M\rangle,
\end{align}
where the non-diagonal operator
\begin{align}
&\hat{\Omega}_\varphi (x_j)= \frac{1}{{8i\rho\gamma}}|\hat{E}^\varphi|^{1/4}\big[\widehat{{\rm sgn}(E^\varphi)}\big(\hat N^\varphi_{2\rho}-\hat N^\varphi_{-2\rho}\big)+\big(\hat N^\varphi_{2\rho}-\hat N^\varphi_{-2\rho}\big)\widehat{{\rm sgn}(E^\varphi)}\big]|\hat{E}^\varphi|^{1/4}\Big|_{x_j},
\end{align}
has a well-defined action on the subspace of states $|\mu_j\rangle$, with  
\begin{equation}
|\hat{E}^\varphi(x_j) |^{1/n} |\mu_j\rangle
= (\gamma\ell_{\rm Pl}^{2}  |\mu_j|)^{1/n}
|\mu_j\rangle,
\end{equation}
for some integer $n$, and
\begin{equation}
\widehat{{\rm sgn}\big(E^\varphi(x_j)\big)}  |\mu_j\rangle
= {\rm sgn}(\mu_j)
|\mu_j\rangle,
\end{equation}
constructed out of the spectral decomposition of $\hat{E}^\varphi$ on
${\cal H}_{\rm kin}$. Besides, we introduce the regularized inverse of the triad (adopting  Thiemann's trick \cite{inv-thiem,inv-thiem1})
\begin{align}
&\widehat{\left[\frac{\left[({E}^x)'\right]^2}{{E}^\varphi}\right]}_{x_j} |\mu_j\rangle
=\sum_{j}^n \delta(x-x_j)\gamma\ell_{\rm Pl}^{2}(\Delta k_j)^2b^2_{\rho_j}(\mu_j)
|\mu_j\rangle.
\end{align}
where 
\begin{equation}
\Delta k_j=k_j-k_{j-1},
\end{equation}
and
\begin{equation}\label{eq:bmu}
b_{\rho_j}(\mu_j)=\frac{1}{\rho_j}(|\mu_j+\rho_j|^{1/2}-|\mu_j-\rho_j|^{1/2}).
\end{equation}
Another useful choice for this regularized inverse triad function is to take the limit $\rho_j\to0$ that formally would correspond to $b_0(\mu_j)=(\mu_j)^{-1/2}$ if $\mu_j\neq 0$, and $b_0(0)=0$.

In total, the action of the constraint on the basis of spin networks is
\begin{align}
&  \hat{H}(N) |\vec{k},\vec{\mu},M\rangle=\frac{1}{G}\sum_{j}^n N(x_j)  \hat C_j|\vec{k},\vec{\mu},M\rangle.
\end{align}
If we recall the decomposition in Eq. \eqref{eq:ket-dec} and the definition of the previous operators, the action of the constraint $\hat C_j$ is given by
\begin{align}\nonumber 
& \hat C_j|\mu_j\rangle=f_0(\mu_j,k_j,k_{j-1},M) |\mu_j\rangle\\
&-f_+(\mu_j)|\mu_j+4\rho_j\rangle-f_-(\mu_j)|\mu_j-4\rho_j\rangle,
\end{align}
with
	\begin{align}\label{eq:f0}
	&f_\pm(\mu_j)=\frac{\ell_{\rm Pl}^2}{64\gamma\rho_j^2}|\mu_j|^{1/4}|\mu_j\pm 2\rho_j|^{1/2}|\mu_j\pm 4\rho_j|^{1/4}s_{\pm}(\mu_j)s_{\pm}(\mu_j\pm 2\rho_j),\\\nonumber
	&f_0(\mu_j,k_j,k_{j-1},M)=\frac{\ell_{\rm Pl}^2}{64\gamma\rho_j^2}\left[(|\mu_j||\mu_j+ 2\rho_j|)^{1/2}s_+(\mu_j)s_-(\mu_j+2\rho_j)
	\right.\\\nonumber
	&\left.+(|\mu_j||\mu_j- 2\rho_j|)^{1/2}s_-(\mu_j)s_+(\mu_j-2\rho_j)\right]+\gamma \ell_{\rm Pl}^2\mu_j\left(1-\frac{2GM}{\sqrt{\gamma\ell_{\rm Pl}^2|k_j|}}\right)\\\label{eq:fpm}
	&-\frac{\gamma\ell_{\rm Pl}^2}{4}(\Delta k_j)^2b^2_{\rho_j}(\mu_j),
	\end{align}

We conclude that the constraint preserves the number of vertices and/or edges, it only relates states with support on semilattices of the form $\mathcal{L}_{\varepsilon_{j}}=\{\mu_{j} | \mu_{j}=\varepsilon_{j}+4\rho_jm,\ m\in \mathbb{N},\ \varepsilon_{j}\in(0,4\rho_j]\}$, and preserves the sequences $\{k_j\}$. Any solution to the scalar constraint will incorporate these properties. 

\section{Solutions to the Hamiltonian constraint}\label{sec:solut}

Any state $(\Psi_g|$ annihilated by the constraint is determined by the condition 
\begin{equation}\label{eq:sol-cond}
\sum_{j}^n(\Phi_g|N(x_j)\hat{C_j}^\dagger=0,\quad\Leftrightarrow \quad (\Phi_g|\hat{C_j}^\dagger=0.
\end{equation}
Since they are well-defined elements in the dual of a dense subspace of the kinematical Hilbert space, they must be of the form 
\begin{equation}
(\Phi_g|=\int_{0}^\infty \!\!dM\sum_{\vec k}\sum_{\vec \mu}\langle \vec{k},\vec{\mu},M|\phi(\vec k,\vec{\mu},M).
\end{equation}
Besides, due to Eq. \eqref{eq:sol-cond}, it is then natural to adopt the factorization 
\begin{equation}
\phi(\vec k,\vec{\mu},M)=\prod_{j}^n\phi_j(\mu_j), \quad \phi_j(\mu_j)=\phi_j(k_j,k_{j-1},\mu_j,M).
\end{equation}
It is possible to see that each $\phi_j(\mu_j)$ fulfills the difference equation
\begin{align}\nonumber
&-f_+(\mu_j-4\rho_j)\phi_j(\mu_j-4\rho_j)-f_-(\mu_j+4\rho_j)\phi_j(\mu_j+4\rho_j)\\
&+f_0(k_j,k_{j-1},\mu_j,M) \phi_j(\mu_j)=0,
\end{align}
with $f_0(k_j,k_{j-1},\mu_j,M)$ and $f_\pm(\mu_j)$ defined in Eqs. \eqref{eq:f0} and \eqref{eq:fpm} , respectively. 

Several properties of the solutions to this set of difference equations is codified in their asymptotic limit. As it was noticed in Ref. \cite{gop}, the asymptotic properties of the solutions depend on the sign of the function
\begin{equation}\label{eq:F}
F_j=\left(1-\frac{2GM}{\sqrt{\gamma\ell_{\rm Pl}^2|k_j|}}\right).
\end{equation}
This quantity allows us to locate on the spin network the quantum analog of the black hole horizon of the classical theory. Although we will not discuss in detail isolated horizons in spherically-symmetric geometries, it is worth commenting that, for those states reproducing smooth semiclassical geometries, the classical considerations for the definition of isolated horizons can be used. However, one must keep in mind that these considerations will only be valid whenever the semiclassical approximation is valid. In more general situations, a more detailed analysis will be required.

As we will explain below, if $F_j>0$, the solutions converge sufficiently fast such that they are normalizable in the kinematical Hilbert space. On the other hand, if $F_j<0$ the solutions are bounded and normalizable in the generalized sense.

\subsection{Solutions for $F_j<0$}

Let us consider those vertices where $k_j$ and $M$ are such that $F_j<0$. In order to deal with the solutions to the constraint, it will be more convenient to write it in a separable form by studying, instead, the solutions 
\begin{equation}
\phi^{\rm cnt}_j(\mu_j)=|\hat{E}^\varphi(x_j) |^{1/2}\phi_j(\mu_j),
\end{equation}
that are related to the original solutions $\phi_j(\mu_j)$ by means of a well-defined bijection (let us recall that $\mu_j=0$ is not included in our analysis). The states $\phi^{\rm cnt}_j(\mu_j)$ are solutions to
\begin{equation}\label{eq:in-eigen-eq}
|\hat{E}^\varphi(x_j) |^{-1/2}\hat C_j|\hat{E}^\varphi(x_j) |^{-1/2}=\hat{\cal C}_j^{\rm cnt}+\left(1-\frac{2GM}{\sqrt{\gamma\ell_{\rm Pl}^2|k_j|}}\right),
\end{equation}
where $\hat{\cal C}_j^{\rm cnt}$ is a difference operator for each vertex. 
It is of the form
\begin{align}\nonumber
&  \hat{\cal C}_j^{\rm cnt} |\mu_j\rangle=
f^{\rm cnt}_0(\mu_j,k_j,M) |\mu_j\rangle
-f^{\rm cnt}_+(\mu_j)|\mu_j+4\rho_j\rangle-f^{\rm cnt}_-(\mu_j)|\mu_j-4\rho_j\rangle,
\end{align}
with 
\begin{align}
&f^{\rm cnt}_\pm(\mu_j)=\frac{1}{64\gamma^2\rho_j^2}|\mu_j|^{-1/4}|\mu_j\pm 2\rho_j|^{1/2}|\mu_j\pm 4\rho_j|^{-1/4}s_{\pm}(\mu_j)s_{\pm}(\mu_j\pm 2\rho_j),\\\nonumber
&f^{\rm cnt}_0(\mu_j,k_j,k_{j-1},M)=\frac{1}{64\gamma^2\rho_j^2|\mu_j|}\left[({|\mu_j||\mu_j+ 2\rho_j|)^{1/2}}s_+(\mu_j)s_-(\mu_j+2\rho_j)\right.\\
&\left.+({|\mu_j||\mu_j- 2\rho_j|)^{1/2}}s_-(\mu_j)s_+(\mu_j-2\rho_j)\right]-\frac{1}{4|\mu_j|}\Delta k_j^2b^2_{\rho_j}(\mu_j).
\end{align}
We can determine its eigenvalues and the corresponding eigenfunctions by means of the difference equation
\begin{equation}
\hat{\cal C}_j^{\rm cnt}|\phi^{\rm cnt}_{\omega_j}\rangle=\omega_j|\phi^{\rm cnt}_{\omega_j}\rangle,
\end{equation}
In this case, if we write $|\phi^{\rm cnt}_{\omega_j}\rangle=\sum_{\mu_j}\phi^{\rm cnt}_{\omega_j}(\mu_j)|\mu_j\rangle$, the equation for the coefficients $\phi^{\rm cnt}_{\omega_j}(\mu_j)$ is 
\begin{equation}\label{eq:cont-sol}
f^{\rm cnt}_0(\mu_j,k_j,M) \phi^{\rm cnt}_{\omega_j}(\mu_j)
-f^{\rm cnt}_+(\mu_j-4\rho_j)\phi^{\rm cnt}_{\omega_j}(\mu_j-4\rho_j)-f^{\rm in}_-(\mu_j+4\rho_j)\phi^{\rm cnt}_{\omega_j}(\mu_j+4\rho_j)=\omega_j\phi^{\rm cnt}_{\omega_j}(\mu_j),
\end{equation}
Let us restrict the study to one of the semilattices $\mathcal{L}_{\varepsilon_{j}}$. They are completely determined up to a normalization constant. Consequently, the eigenvalues are nondegenerate (whenever all the other quantum numbers like $\{k_j\}$ and $M$ have been fixed). We will assume that $\phi^{\rm cnt}_{\omega_j}(\mu_j)$ is a smooth function of $\mu_j$ for $\mu_j\gg\rho_j$. The difference equation (times $\gamma^2$) is well approximated  by 
\begin{equation}
-\partial^2_{\mu_j}\phi^{\rm cnt}_{\omega_j}(\mu_j)-\frac{3+4\gamma^2(\Delta k_j)^2}{16\mu_j^{2}}\phi^{\rm cnt}_{\omega_j}(\mu_j)=\gamma^2\omega_j\phi^{\rm cnt}_{\omega_j}(\mu_j).
\end{equation}
Therefore, any solution $\phi^{\rm cnt}_{\omega_j}(\mu_j)$ must correspond to a linear combination of the solutions to this differential equation, i.e.,
\begin{equation}
\phi^{\rm cnt}_{\omega_j}(\mu_j)\stackrel{\mu_j\gg\rho_j}{\simeq}A\sqrt{\mu_j}H^{(1)}_{i\kappa}\left(\sqrt{\omega_j}\gamma\mu_j\right)+B\sqrt{\mu_j}H^{(2)}_{i\kappa}\left(\sqrt{\omega_j}\gamma\mu_j\right),
\end{equation}
where $H^{(1)}_{\nu}(x)$ and $H^{(2)}_{\nu}(x)$ are the Hankel functions of first and second kind, respectively, $16\kappa^2=4\gamma^2(\Delta k_j)^2-1$.
Therefore, one can infer easily that for $\omega_j\geq 0$ the solutions have a oscillatory behavior when $\mu_j\gg\rho_j$, irrespective of the value of $\Delta k_j$. 

However, this is only valid whenever the functions $\phi^{\rm cnt}_{\omega_j}(\mu_j)$ are smooth, and this only happens for some values of $\omega_j$. Indeed, there are values of $\omega_j$ above an upper bound where no normalizable eigenfunctions have been found (although it is possible that some of them exist). In order to find this bound, one can follow the analysis of Ref. \cite{scatter} (see Ref. \cite{gowdy-LRS} for its application to a similar model). We will not show here the calculation, but it is possible to prove that $\omega_j\in[0,(2\gamma \rho_j)^{-2}]$. This upper bound in the spectrum of this operator is a genuine consequence of the loop quantization. It is not completely clear to us which is its physical interpretation, but we believe that it is related with the quantum bounce in Kantowski---Sachs models \cite{cgp2,ash-bojo}.

The eigenfunctions fulfill the normalization condition 
\begin{equation}
\langle \phi^{\rm in}_{\omega_j}|\phi^{\rm in}_{\omega'_j}\rangle=\delta\left(\sqrt{\omega_j}-\sqrt{\omega_j'}\right),
\end{equation}
for $\omega_j\in[0,(2\gamma \rho_j)^{-2}]$. Although for negative eigenvalues, as well as complex ones, we have not analyzed the difference equation with sufficient detail, they are not expected to be present or at least not to contribute to the final physical description. We will then disregard them.

The constraint equation takes the diagonal form
\begin{equation}\label{eq:const-eq}
\omega_j+\left(1-\frac{2GM}{\sqrt{\gamma\ell_{\rm Pl}^2|k_j|}}\right)=0,
\end{equation}
in this basis.

\subsection{Solutions for $F_j>0$}

We will continue here with those vertices where $k_j$ and $M$ take values such that $F_j>0$. We will consider the solutions
\begin{equation}\label{eq:scaling-out}
\phi^{\rm dcr}_j(\mu_j)=\frac{1}{\gamma \ell^2_{\rm Pl}}\hat b_{\rho_j}(\mu_j)\phi_j(\mu_j),
\end{equation}
where  $b_{\rho_j}(\mu_j)$ is given in \eqref{eq:bmu}, of a new difference equation that takes a separable form. For a vertex $v_j$, it takes the form
\begin{equation}\label{eq:const-ext}
\left(\frac{1}{\gamma \ell^2_{\rm Pl}}\hat b_{\rho_j}(\mu_j)\right)^{-1}\hat C_j\left(\frac{1}{\gamma \ell^2_{\rm Pl}}\hat b_{\rho_j}(\mu_j)\right)^{-1}=\hat{\cal C}_j^{\rm dcr}-\frac{1}{4}(\Delta k_j)^2,
\end{equation}
with $\hat{\cal C}_j^{\rm dcr}$ a difference operator whose spectral decomposition can be carried out. The action of this operator $\hat{\cal C}_j^{\rm dcr}$ on a kinematical state is
\begin{align}
&  \hat{\cal C}_j^{\rm dcr} |\mu_j\rangle=
f^{\rm dcr}_0(\mu_j,k_j,M) |\mu_j\rangle
-f^{\rm dcr}_+(\mu_j)|\mu_j+4\rho_j\rangle-f^{\rm dcr}_-(\mu_j)|\mu_j-4\rho_j\rangle,
\end{align}
with 
\begin{align}
&f^{\rm dcr}_\pm(\mu_j)=\frac{1}{64\gamma^2\rho_j^2}\frac{|\mu_j|^{1/4}|\mu_j\pm 2\rho_j|^{1/2}|\mu_j\pm 4\rho_j|^{1/4}s_{\pm}(\mu_j)s_{\pm}(\mu_j\pm 2\rho_j)}{b_{\rho_j}(\mu_j)b_{\rho_j}(\mu_j\pm 4\rho_j)},\\\nonumber
&f^{\rm dcr}_0(\mu_j,k_j,k_{j-1},M)=\frac{\mu_j}{b_{\rho_j}(\mu_j)^2}\left(1-\frac{2GM}{\sqrt{\gamma\ell_{\rm Pl}^2|k_j|}}\right)+\frac{1}{64\gamma^2\rho_j^2}\left[\frac{({|\mu_j||\mu_j+ 2\rho_j|)^{1/2}}s_+(\mu_j)s_-(\mu_j+2\rho_j)}{b_{\rho_j}(\mu_j)^2}\right.\\
&\left.+\frac{({|\mu_j||\mu_j-2 \rho_j|)^{1/2}}s_-(\mu_j)s_+(\mu_j-2\rho_j)}{b_{\rho_j}(\mu_j)^2}\right].
\end{align}
We will study its positive spectrum by solving the eigenvalue problem
\begin{equation}\label{eq:dcr-eigsys}
\hat{\cal C}_j^{\rm dcr}|\phi^{\rm dcr}_{\lambda_j}\rangle=\lambda_j|\phi^{\rm dcr}_{\lambda_j}\rangle.
\end{equation}

If the solutions are assumed to be of the form $|\phi^{\rm dcr}_{\lambda_j}\rangle=\sum_{\mu_j}\phi^{\rm dcr}_{\lambda_j}(\mu_j)|\mu_j\rangle$, the Eq. \eqref{eq:dcr-eigsys} becomes the difference equation for the coefficients $\phi^{\rm dcr}_{\lambda_j}(\mu_j)$ 
\begin{align}\label{eq:difference-eq}\nonumber
&-f^{\rm dcr}_+(\mu_j-4\rho_j)\phi_{\lambda_j}^{\rm dcr}(\mu_j-4\rho_j)-f^{\rm dcr}_-(\mu_j+4\rho_j)\phi_{\lambda_j}^{\rm dcr}(\mu_j+4\rho_j)\\
&+f^{\rm dcr}_0(k_j,k_{j-1},\mu_j,M) \phi_{\lambda_j}^{\rm dcr}(\mu_j)=\lambda_j \phi_{\lambda_j}^{\rm dcr}(\mu_j).
\end{align}

The eigenfunctions will have support on semilattices $\mathcal{L}_{\varepsilon_{j}}$. Let us restrict the study to one of them. The eigenfunctions are determined up to normalization. The spectrum is then nondegenerate.  Again, we can extract several properties of the eigenfunctions in the regime $\mu_j\gg \rho_j$, assuming they are smooth there. In that regime, the solutions to the difference equation (up to a global $\gamma$ factor) can be approximated by the differential one
\begin{equation}\label{eq:mod-bessel-diff-op}
-\mu_j^2\partial_{\mu_j}^2\phi_{\lambda_j}^{\rm dcr}(\mu_j)-2\mu_j\partial_{\mu_j}\phi_{\lambda_j}^{\rm dcr}(\mu_j) +\mu_j^2\gamma^2F_j\phi_{\lambda_j}^{\rm dcr}(\mu_j)=\left(\gamma^2\lambda_j+\frac{3}{16}\right)\phi_{\lambda_j}^{\rm dcr}(\mu_j).
\end{equation}
The solutions to this differential equation are linear combinations of modified Bessel
functions, i.e.,
\begin{equation}
\phi^{\rm dcr}_{\lambda_j}(\mu_j)\stackrel{\mu_j\gg\rho_j}{\simeq}A x_j^{-1/2}{\cal K}_{-i\kappa_j}\left(x_j\right)+B x_j^{-1/2}{\cal I}_{-i\kappa_j}\left(x_j\right),
\end{equation}
with $x_j=\mu_j\gamma\sqrt{ F_j}$ and $\kappa_j^2=\gamma^2\lambda_j-\frac{1}{16}$. In the limit $\mu_j\to\infty$, ${\cal I}$ grows exponentially, and ${\cal K}$ decays exponentially. Therefore, the latter is the only contribution to the spectral decomposition. As a consequence, this
counterpart of the spectrum of the differential operator \eqref{eq:mod-bessel-diff-op} is nondegenerate (as before for a given choice of $M$ and $\{k_j\}$). Moreover, the functions
${\cal K}_{i\kappa}(x)$ are normalized to
\begin{equation}
\langle {\cal K}_{i\kappa}|{\cal K}_{i\kappa '}\rangle=\delta(\kappa-\kappa'),
\end{equation}
in $L^2(\mathbb{R},x^{-1}dx)$, since the normalization in this case is ruled by the behavior of
${\cal K}_{i\kappa}(x)$ in the limit $x\to 0$, which corresponds to
\begin{equation}
\lim_{x\to 0}{\cal K}_{i\kappa}(x)\to A\cos\left(\kappa\ln|x|\right).
\end{equation}
For additional details, see also Ref.~\cite{kiefer}. This normalization condition is valid if and only if $\lambda_j\geq\frac{1}{16\gamma^2}$. Otherwise, the solutions that are normalizable at $x\to \infty$ diverge at $x\to 0$ and vice versa. This fact is important for the solutions to the differential equation, but for the ones to Eq. \eqref{eq:difference-eq} there is no such restriction. There, the limit $x\to 0$ cannot be taken since $\mu_j\geq \varepsilon_j>0$ at any vertex $x_j$. So, we only need that the eigenfunctions be normalizable for $\mu_j\to\infty$ since they are bounded and well defined anywhere else.

We obtain the eigenfunctions following the ideas of Ref.~\cite{cfrw} (where the homogeneous constraint equation is analogous to ours at each vertex). 
The spectrum of the corresponding difference operator turns out to be discrete due to the behavior of its eigenfunctions at $\mu_j\sim \varepsilon_j$ and $\mu_j\to \infty$. At $\mu_j\sim \varepsilon_j$, the eigenfunctions cannot oscillate infinitely before reaching $\varepsilon_j$, and for some values $\lambda_j$, they will decrease exponentially at $\mu_j\to \infty$. Therefore, we expect that $\lambda_j$ will belong to a countable set, which can be determined numerically. We will define the elements of this set as $\lambda_p(\varepsilon_j)$, for some integer $p$, since our preliminary numerical investigations show that it depends on $\varepsilon_j\in(0,4\rho_j]$, as well as on $F_j$, i.e., on $k_j$ and $M$ by means of Eq.~\eqref{eq:F}.

The corresponding eigenfunctions are normalized to
\begin{equation}
\langle \phi^{\rm out}_{\lambda_p(\varepsilon_j)}|\phi^{\rm out}_{\lambda_{p'}(\varepsilon_j')}\rangle=\delta_{pp'}\delta_{jj'}.
\end{equation}

Therefore, the constraint equation in this basis takes the algebraic form
\begin{equation}\label{eq:algeb-const}
\lambda_p(\varepsilon_j)-\frac{1}{4}(\Delta k_j)^2=0.
\end{equation}
At this point, it is interesting to ask which are the physical consequences if Eq. \eqref{eq:algeb-const} must be fulfilled. Let us notice that there seems to be some tension since the second term will be the square of an integer (over four), while $\lambda_p(\varepsilon_j)$ is discrete once $F_j$ and $\varepsilon_j$ are fixed. In order to alleviate this tension, we can consider two strategies. The first one is to choose a value for $F_j$ and then check the dependence of the eigenvalues with respect to the parameter $\varepsilon_j$, that can take any real value in $(0,4\rho_j]$, and see whether it is possible to find a suitable choice of $p$ and $\varepsilon_j$  such that Eq. \eqref{eq:algeb-const} is fulfilled. This is feasible as we have seen from our numerical studies. The physical consequence is that physical states will be superpositions of states on different semilattices. Therefore, the discretization of the triad $\hat E^\varphi$ will become smoother. On the other hand, the second possibility is to restrict the numerical study to one semilattice given by $\varepsilon_j$ and to check which values of $F_j$ are compatible with Eq. \eqref{eq:algeb-const}. This possibility seems to be too restrictive. In the asymptotic region $k_j\to\infty$, where $F_j\simeq 1$, the possible values of $\Delta k_j$ seem to be strongly restricted. Therefore, the most plausible situation for solving the mentioned tension is to allow for superpositions of different $\varepsilon_j$.

\section{Physical Hilbert space and observables}\label{sec:physical}

In order to determine the physical states and a suitable inner product, we will carry out a group averaging \cite{raq} of the kinematical states with respect to the spacetime diffeomorphisms. The spatial diffeomorphisms and its implementation in the quantum theory are well known \cite{bbr,raq-lqg}. However, this is not the case for the transformations generated by the quantum scalar constraint. This is why we define its generator at the quantum level, and then we group average the kinematical states. Since we have been able to diagonalize it at any vertex, the calculation can be carried out straightforwardly \cite{gop,gop2}. The resulting states after averaging will be invariant under the spacetime transformations. 

We will not show here the calculation, but one can see that the solutions to the scalar constraint take the form
\begin{equation}\label{eq:solution}
(\Psi^{C}_g|=\int_0^{\infty} dM\left(\bigotimes_j\left[\sum_{k_j}\psi(M,k_j)\langle\phi(k_j,M)|\otimes\langle k_j| \right]\right)\otimes \langle M|,
\end{equation}
where $\langle \phi(\vec{k},M) |$ means that each eigenstate described in the previous section are such that the corresponding eigenvalue satisfies either Eq. \eqref{eq:const-eq} or Eq. \eqref{eq:algeb-const} (see Refs. \cite{gop,gop2}). 

In addition, we must provide the (spatially) diffeomorphism-invariant states in order to complete the quantization program. They are well known in the full theory, as well as in this particular reduced model.\footnote{ See Appendix A of Ref. \cite{bbr} for a recent discussion about this issue.} The strategy we adopt here is to carry out group averaging of the previous solutions to the Hamiltonian constraint with respect to the group of spatial diffeomorphism \cite{raq-lqg}. A spatial diffeomorphism produces a dragging of the vertices of the corresponding state preserving their order and valences. In order to construct a physical state that is invariant under such transformations, roughly speaking, we must consider an arbitrary solution to the Hamiltonian constraint and sum all possible states that are related with the reference one by a spatial diffeomorphism. This can be carried out since the group of spatial diffeomorphisms is well known: it is the group of invertible diffeomorphisms on a one-dimensional open manifold. 

This construction yields a well-defined inner product 
\begin{equation}\label{eq:undep-inner}
\|\Psi_{\rm Phys}\|^2=(\Psi_{\rm Phys}|\Psi_{\rm Kin}\rangle=\int_0^{\infty} dM\sum_{\vec{k}}|\psi(M,\vec{k})|^2.
\end{equation}

This physical picture must be completed with suitable observables acting on physical states. If we look at the previous inner product, there is a natural basis of physical states $|M,\vec k\rangle_{\rm Phys}$, where the basic observables are given by $\hat M$, the mass of the black hole and a new observable without classical Dirac analogue, defined as
\begin{equation}\label{eq:ObsO}
\hat O(z)|M,\vec k\rangle_{\rm Phys}=\gamma\ell_{\rm Pl}^2 k_{{\rm Int}(nz)}|M,\vec k\rangle_{\rm Phys},
\end{equation}
recalling that $n$ is the number of vertices, and with $z\in [0, 1]$ a gauge parameter, and ${\rm Int}(nz)$ means the integer part of $n z$. It codifies the sequence of the areas of the spheres of symmetry (which are quantized). 

With this set of basic observables, one can promote kinematical operators (that are not gauge invariant) to parametrized observables (operators that only depend on suitable gauge parameters and observables). This is the strategy discussed, for instance, in Refs. \cite{gp-lett,gop,gop2,mgp,cgop,gp-semi}. 

For convenience, we will instead present here the ideas of Ref. \cite{gop2}, where we parametrize the states, instead of the observables. This is possible since the inner product in Eq. \eqref{eq:undep-inner} involves a kinematical state $|\Psi\rangle_{\rm kin}$, any of them defined on a given one-dimensional manifold with the vertices located at concrete positions on this manifold, in such a way that in Eq. \eqref{eq:undep-inner} only the projection of $\langle \Psi_{\rm Phys}|$ on $|\Psi\rangle_{\rm kin}$ will contribute. In addition, different choices of constant time surfaces correspond to the freedom existing in order to select the relational time in the system\footnote{For instance, in analogy with quantum cosmology, one declares a suitable phase space function as an internal time (gauge parameter). In this way the physical system is parametrized in terms of this physical clock.}.

We will then proceed as follows. Since we can characterize all the kinematical states related to $|\Psi\rangle_{\rm kin}$ by means of a spatial diffeomorphism $x\to z(x)$ as $|\Psi\rangle_{\rm kin}\to|\Psi\left(z\right)\rangle_{\rm kin}$, we can define analogously the family of all physical states projected on these kinematical states as $\langle\Psi_{\rm Phys}\left(z\right)|$ (notice that $\langle\Psi_{\rm Phys}|$ can be understood as the sum in $z$ of the states $\langle\Psi_{\rm Phys}\left(z\right)|$). This gives a suitable parametrization with respect to the choice of spatial coordinates for the physical states. In addition,  we will parametrize the physical states for a particular choice of time function. For simplicity, we will choose the connection $K_\varphi(x)$ as our relational time.\footnote{Although it is a well defined time function, as it has already been noticed in loop quantum cosmology, for instance, in Refs. \cite{bianchi1,cgp2}, from a physical point of view is not accessible since it is not well defined as an operator in the quantum theory.} We then parametrize the physical states, such that
\begin{align}\label{eq:solution-dep}\nonumber
&\langle\Psi_{\rm Phys}\left(z,\vec \eta^{(0)}\right)|=\sqrt{2\pi} \langle\Psi_{\rm Phys}\left(z\right)|\vec K_\varphi=\vec \eta^{(0)}\rangle\\
&=\int_0^{\infty} dM\left(\bigotimes_{z(v_j)}\left[\sum_{k_j}\sum_{\mu_j}\psi(M,k_j)\phi(k_j,M;\mu_j)e^{i\gamma \ell_{\rm Pl}^2 \mu_j\eta^{(0)}_j}\langle k_j| \right]\right)\otimes \langle M|.
\end{align} 
Here, $\eta_j=K_\varphi(x_j)$ is the collection of parameters associated with the values of the connection $K_\varphi(x)$ evaluated on to the vertices of $|\Psi_{\rm Phys}\left(z\right)\rangle$. Besides, we also declare $\{\eta^{(0)}_j\}$ as our ``initial'' Cauchy surface. 

The reader must realize that these bras provide a natural definition of the corresponding kets $|\Psi_{\rm Phys}\left(z,\vec \eta^{(0)}\right)\rangle$. It is important to notice that the inner product defined above in Eq. \eqref{eq:undep-inner} coincides with $\langle\Psi_{\rm Phys}\left(z,\vec \eta^{(0)}\right)|\Psi_{\rm Phys}\left(z,\vec \eta^{(0)}\right)\rangle$, where the two parametrized physical states must correspond to the same choice of $z(x)$ and Cauchy surface $\{\eta^{(0)}_j\}$. More explicitly,
\begin{equation}
\|\Psi_{\rm Phys}\|^2=\langle\Psi_{\rm Phys}\left(z,\vec \eta^{(0)}\right)|\Psi_{\rm Phys}\left(z,\vec \eta^{(0)}\right)\rangle.
\end{equation}

Besides, there is a map 
\begin{equation}
\hat U\left(\vec \eta,\vec \eta^{(0)}\right): \quad |\Psi_{\rm Phys}\left(z,\vec \eta^{(0)}\right)\rangle\to |\Psi_{\rm Phys}\left(z,\vec \eta\right) \rangle = \hat U\left(\vec \eta,\vec \eta^{(0)}\right)|\Psi_{\rm Phys}\left(z,\vec \eta^{(0)}\right)\rangle,
\end{equation}
that relates the parametrized physical states between different slicings $\{\eta^{(0)}_j\}$ and $\{\eta_j\}$. It is the analog to the evolution operator in quantum mechanics in the Schr\"odinger picture. It admits a decomposition as the product of evolution operators acting on each vertex separately, i.e.,
\begin{equation}
\hat U(\vec \eta,\vec \eta^{(0)})=\prod_{j}^n \hat U_j(\eta_j,\eta^{(0)}_j).
\end{equation}

These operators can be constructed explicitly in the $\mu_j$-representation simply as
\begin{equation}
\hat U_j(\eta_j,\eta^{(0)}_j)|\mu_j\rangle:= \exp\left\{i\gamma \ell^2_{\rm Pl}\mu_j(\eta_j-\eta^{(0)}_j)\right\}|\mu_j\rangle.
\end{equation}
Then, the mass and triad operators can be defined easily as physical relational observables as follows
\begin{subequations}
	\begin{align}\label{eq:solution-dep-Ex}
	&\hat M|\Psi_{\rm Phys}\left(z,\vec \eta^{(0)}\right)\rangle=\\\nonumber
	&\int_0^{\infty} dM\left(\bigotimes_{z(v_j)}\left[\sum_{k_j}\sum_{\mu_j}M\psi(M,k_j)\phi(k_j,M;\mu_j)e^{i\gamma\ell_{\rm Pl}^2 \mu_j\eta^{(0)}_j}|k_j\rangle \right]\right)\otimes |M\rangle.\\
	&\hat E^x(x)|\Psi_{\rm Phys}\left(z,\vec \eta^{(0)}\right)\rangle=\\\nonumber
	&\int_0^{\infty} dM\left(\bigotimes_{z(v_j)}\left[\sum_{k_j}\sum_{\mu_j}\gamma\ell_{\rm Pl}^2 k_{{\rm Int}(Nz)}\psi(M,k_j)\phi(k_j,M;\mu_j)e^{i\gamma\ell_{\rm Pl}^2 \mu_j\eta^{(0)}_j}|k_j\rangle \right]\right)\otimes |M\rangle.\\
	&\hat E^\varphi(x)|\Psi_{\rm Phys}\left(z,\vec \eta^{(0)}\right)\rangle=\\\nonumber
	&\int_0^{\infty} dM\left(\bigotimes_{z(v_j)}\left[\sum_{k_j}\sum_{\mu_j}\gamma\ell_{\rm Pl}^2 \mu_{{\rm Int}(Nz)}\psi(M,k_j)\phi(k_j,M;\mu_j)e^{i\gamma\ell_{\rm Pl}^2 \mu_j\eta^{(0)}_j}|k_j\rangle \right]\right)\otimes |M\rangle.
	\end{align} 
\end{subequations}

As we have seen, kinematical operators are promoted to physical observables in a simple way. It is not necessary to know a priori the parametrized observables in this picture. However, this is possible since we have been able to apply group averaging techniques. In those situations where they cannot be used, the knowledge of the (parametrized) observables is a key ingredient in order to complete the quantization, as we will see in the next sections.

\section{Comparison with previous quantizations}\label{sec:comp}

In this section, we will briefly explain previous approaches for the quantization of spherically-symmetric spacetimes. All of them are based on further symmetry reduction or partial gauge fixing. For instance, the interior of the black hole can be identified with a Kantowski---Suchs spacetime after a symmetry reduction (requiring homogeneity) \cite{ash-bojo} or adopting a gauge fixing \cite{cgp2}. The former proposal adopts a quantization prescription that is closely related to the one of the full theory \cite{lqg} but at the same time more difficult to solve. The quantization proposal of Ref. \cite{cgp2} uses a quantization prescription where it is possible to construct explicitly the solutions to the quantum scalar constraint and to show that the classical singularity is naturally resolved. Besides, if one instead considers a quantization for the exterior of the black adopting a partial gauge fixing, the resulting reduced theory can be quantized as well within a loop representation and the solutions to the constraint determined. We will give further details below.  

\subsection{Interior of the black hole: Kantowski-Suchs spacetimes}

In order to study the interior of the black hole, we follow the analysis of Ref. \cite{cgp2}. There, one considers a vacuum spherically-symmetric spacetime, like the one provided in Sec. \ref{sec:class-syst}. Then, one introduces at the classical level the second class condition $\phi= (E^x)'=0$. 
Preservation of this condition in time, together with the diffeomorphism constraint and $\phi=0$, determines partially the lapse, but introduces an additional second class condition $K_\varphi'=0$. It must be preserved upon evolution, which implies $(E^\varphi)'=0$. The preservation of the previous conditions in time guaranties that this last condition is preserved, as well. If in addition, $K_x$ is chosen initially independent of the radial coordinate, all kinematical variables become independent of it. 

One is left with only one constraint, the homogeneous Hamiltonian constraint. Before writing it explicitly, it will be convenient to introduce the variables that are commonly used in loop quantum cosmology. The new variables
are related with the previous ones by $b=\gamma K_\varphi$, $c=\gamma K_x L_0$,
$p_b=E^\varphi L_0$, $p_c=E^x$, where $L_0$ is a fiducial length associated with the fiducial radial coordinate in the interior of the black hole. The homogeneous Hamiltonian constraint (up to a factor $G^{-1}$) takes the form,
\begin{equation}\label{hamiltonian}
H = -\frac{p_b}{2} -\frac{2 b\, c\, p_c}{\gamma^2}-\frac{1}{2}\frac{b^2\,
	p_b}{\gamma^2},
\end{equation}
where we introduce here the same densitization by a factor $1/\sqrt{p_c}$ for the constraint of Ref. \cite{gop2}. Notice that a change in $x$ modifies $L_0$, and it affects the value of $p_b$ and $c$. See Ref. \cite{cs} for a recent discussion. 

In the process of quantization, since the connection is not a well-defined operator, but the holonomies of the connection, the quantum scalar constraint must be suitably written in terms of them. There are different choices available in the literature. See, for instance, \cite{ash-bojo,vander,cgp2,modesto,modesto1}. Moreover, there is additional freedom in the particular approach quantization, either within the $\mu$ or the $\bar \mu$ schemes. For simplicity, we will consider here the $\mu$ scheme. For additional details about the other approach, see Refs. \cite{vander,cs}. 

If one polymerizes the Hamiltonian constraint following the ideas of \cite{cgp2}, and rearrange it at the classical level, upon quantization in the $(b,c)$-representation, a symmetric Hamiltonian constraint operator turns out to be
\begin{eqnarray}\label{26}
\hat{{\bf H}}&=&-i \ell_{\rm Pl}^2 \gamma\frac{\partial }
{\partial c} -\frac{i \ell_{\rm Pl}^2 \gamma}
{4\sin(\mu\,c)\sin(\mu\,b)} \left[
\left(\sin(\mu\,b)^2+\mu^2\gamma^2\right)
\frac{\partial }
{\partial b}+\frac{\mu\cos(\mu\,b) }{2\sin(\mu\,b)}
\left(\sin(\mu\,b)^2 -\mu^2 \gamma^2\right)
\right].\nonumber
\end{eqnarray}
If one writes the solutions to this constraint in the form
\begin{equation}\label{27}
\Psi\left(b,c\right) = 
A\left(b,c\right) \exp\left( \frac{ik}{\ell_{\rm Pl}^2}
S\left(b,c\right)\right),
\end{equation} 
with $k$ a constant with dimensions of length to the square and with $A(b,c)$ given by,
\begin{equation}\label{abc}
A(b,c) = C_1\sqrt{\frac{\sin^2(\mu\,b)+\gamma^2\mu^2}{\left(1+\gamma^2\mu^2
		\right)|\sin(\mu\,b)|}}
\end{equation}
for any nonvanishing $C_1$, the function $S(b,c)$ fulfills 
\begin{equation}
\left[\frac{\mu^2\gamma^2}{4\sin(\mu\,b)\sin(\mu\,c)}
+\frac{\sin(\mu\, b)}{4\sin(\mu\, c)}\right]
\frac{\partial S(b,c)}{\partial b} +
\frac{\partial S(b,c)}{\partial c}
=0.
\end{equation}
The solutions to this equation are arbitrary function $S(w(b,c))$ with 
\begin{equation}
w(b,c) = \ln\left(\frac{\sin(\mu\,c)}{\cos(\mu\,c)+1}\right)
+ 4 \frac{\tanh^{-1}\left(\frac{\cos(\mu\,b)}{\sqrt{\gamma^2\mu^2+1}}\right)}
{\sqrt{\gamma^2\mu^2+1}}.
\end{equation}

These solutions satisfy the boundary conditions 
$\Psi(0,c)=\Psi(\pi/\mu,c)$,
$\Psi(b,0)=\Psi(b,\pi/\mu)$.  

If we consider $c$ as an internal time to the evolution of the system, we only need to specify an initial Cauchy surface at $c_0$, physical observables acting on the physical Hilbert space and a unitary evolution. 

In Ref. \cite{cgp2}, it was shown that it is possible to define a unitary evolution with respect to the internal time $c$. The corresponding Hamiltonian operator is given by 
\begin{eqnarray}\label{eq:Htrue}
\hat H_{\rm True}=-\frac{i \ell_{\rm Pl}^2 \gamma}
{4\sin(\mu\,c)\sin(\mu\,b)} \left[
\left(\sin(\mu\,b)^2+\mu^2\gamma^2\right)
\frac{\partial }
{\partial b}+\frac{\mu\cos(\mu\,b) }{2\sin(\mu\,b)}
\left(\sin(\mu\,b)^2 -\mu^2 \gamma^2\right)
\right].\nonumber
\end{eqnarray}

Since the dimensionality of the two subspaces defined as ${\cal K}_\pm = {\rm ker}( H_{\rm True}\pm i)$ coincides, where the normalizable states belonging to each kernel are 
\begin{equation}
\tilde \Psi\left(b,c\right) = \Psi\left(b,c\right) \exp\left(\mp4\, \tanh^{-1} \left( {\frac {
		\cos \left( \mu\,{\it b}\right)  \sin \left( \mu\,{\it c} \right)}
	{\gamma \mu \left({{\mu}^{2}{\gamma}^{2}+1}\right)}}\right)\right), 
\end{equation}
one then concludes that $\hat H_{\rm True}$ is self-adjoint, and a unitary evolution with respect to $c$ exists. Besides, since the solutions to the constraint are normalizable with respect to the kinematical inner product, the quantization can be easily completed. Suitable quantum Dirac observables acting on this space can be constructed by means of the kinematical ones as relational observables following, for instance, the ideas of Sec. \ref{sec:physical}.

Although we lack a more detailed study about the quantum dynamics of this model, we can extract the effective semiclassical one of this loop quantization of the interior of the black hole. It is enough to analyze the Eikonal approximation for $S$. For simplicity, we will choose it to be equal to $w(b,c)$. We may also notice that we have not specified yet the parameter $\mu$. 

Following Hamilton--Jacobi theory, we compute the canonical momenta of the original configuration variables, in this case $b$ and $c$, as
\begin{eqnarray}
\label{pb}
p_b &=& \frac{4 k \sin(\mu\,b) \mu}
{\gamma^2\mu^2 +1 -\cos(\mu\,b)^2},\\
p_c &=& -\frac{k \mu}{\sin(\mu\,c)}\label{pc},
\end{eqnarray}
as well as the momentum of the constant of integration, in this case
$k$, which is also a constant of the motion,
\begin{equation}
p_k = -\ln\left(\frac{\sin(\mu\,c)}{\cos(\mu\,c)+1}\right)
-\frac{4 \tanh^{-1}\left(\frac{\cos(\mu\,b)}{\sqrt{\gamma^2\mu^2+1}}\right)}
{\sqrt{\gamma^2\mu^2+1}}.
\end{equation}

If we adopt again $c$ as the time parameter and we can write all the previous quantities as functions of it, the constants of integration $(k,p_k)$ and the initial data $(b_0,p_{b,0})$ and $(c_0,p_{c,0})$. Following the results of Ref. \cite{cgp2}, we will consider a particular set of initial data, but any other choice can be adopted, as well. There, one takes $b_0=0$ and $p_{c,0}=4M^2$. This involves $p_{b,0}=0$ and $c_0$ undetermined. The constants $k$ and $p_k$ can be written in terms of the previous initial data and $M$. Indeed, they can be written in terms of $c$ and $b$ at any time.  These relations evaluated at $(c_0,b_0)$ and another arbitrary choice $(c,b)$ allows one to compute $b$ as a function of $c$ as
\begin{eqnarray}
b &=& \frac{1}{\mu} \cos^{-1}\left[\sqrt{1+\gamma^2\mu^2} \tanh\left(
\tanh^{-1}\left(\frac{1}{\sqrt{1+\gamma^2\mu^2}}\right)\right.\right.\nonumber\\
&&\left.\left. +
\frac{\sqrt{1+\gamma^2\mu^2}}{4}\left(-\ln\left(\tan(\frac{\mu\, c}{2})\right)+
\ln\left(\tan(\frac{\mu\,c_0}{2})\right)\right)\right)\right].
\end{eqnarray}
Then, any other phase space variable can be written as functions of the internal time $c$, in particular $p_b$ and $p_c$, through Eq. \eqref{pb}. 

The resulting expressions are sufficiently complicated that the use of numerical tools was required in Ref. \cite{cgp2}. There, it was shown that the volume of the spacetime evolving as a function of $c$ experiences a bounce replacing the singularity of general relativity. This is in agreement with the singularity resolution discussed in Sec. \ref{sec:scalar-const}. Besides, the studies of Ref. \cite{cs} about the semiclassical dynamics in the $\bar\mu$ scheme show a similar behavior, giving robustness to the fact that loop quantum gravity is able to successfully cure the singularity of a black hole.

\subsection{Exterior of the black hole: Schwarzschild spacetime}

The quantization of the exterior of the black hole can also be carried out within loop quantum gravity. We will follow the main ideas of Ref. \cite{cgp}. The first step is to adopt a suitable gauge fixing condition, like $E^x(x) = (x+2GM)^2$. Imposing this gauge fixing condition classically and solving the diffeomorphism constraint, we have at any $x$
\begin{equation}
K_x(x)=\frac{E^\varphi(x)
	K'_\varphi(x)}{2(x+2GM)}.
\end{equation}
The new Hamiltonian constraint in Eq. \eqref{eq:H_new-den}, after the partial gauge fixing, takes the form
\begin{equation}
\tilde H(\tilde N) :=\frac{1}{G}\int dx \tilde N
E^\varphi\bigg[ K_\varphi^2-\frac{(x+2 G M)^2}{
	(E^\varphi)^2}+\frac{x}{x+2 G M}\bigg].
\end{equation}
The constraint only depends on the canonical pair
$K_\varphi(x),E^\varphi(x)$, $M$, and the radial coordinate $x$. One can see that this constraint has an Abelian algebra with itself.

At the quantum level, we will adopt a kinematical Hilbert space mimicking the one of the full theory. As in Ref. \cite{cgp}, one possible choice is to consider a one-dimensional lattice $L$ with points $0,\ldots, x_j$ distributed on $\mathbb{R}^+$ such that the spacing between two consecutive points is $\epsilon_j=x_{j+1}-x_{j}$. The kinematical Hilbert space is given by
\begin{equation}
{\cal H}_{\rm kin} = L^2(\mathbb{R},dM)\bigotimes_j^n \ell_j^2
\end{equation}
where $\ell_j^2$ means the set of square summable functions associated with the vertices. In this space one
can introduce a basis $|M,\vec{\mu}\rangle$. In the $(\tau,K_\varphi)$-representation,  the elements of the basis take the form
\begin{equation}
\Psi_{M,\vec{\mu}}[\tau,K_\varphi]=e^{iM\tau}\prod_{j}^n 
\exp\left( i\mu_j K_\varphi(x_j)\right)
\end{equation}
with $\mu_j\in \mathcal{L}_{\varepsilon_j}$, such that $\mathcal{L}_{\varepsilon_j}$ is a lattice in $\mathbb{R}$ of constant step $4\rho_j$, for some positive polymer parameters $\rho_j$, and with  $\varepsilon_j$ its smaller positive value. Each lattice can be understood as a collection of vertices. It is worth commenting that it is not necessary to introduce the lattices in order to study the quantization of the model, and a representation in the continuum can also be adopted.

This representation has the advantage that the theory is automatically regularized. For instance, the triad operator takes the simple form
\begin{equation}\label{eq:ephi-kin}
\hat{E}^\varphi(x_j) |M,\vec{\mu}\rangle =\sum_{k} \mu_k
\gamma \ell_{\rm Pl}^2 \delta_{k,j} |M,\vec{\mu}\rangle,
\end{equation}
where $\delta_{k,j}$ is a Kronecker delta (instead of a Dirac one like in Eq. \eqref{eq:evarphi}).

Now, the strategy that we will follow here in order to find the physical states is the one adopted in Refs. \cite{cgp,gp-lett}. After rearranging the classical constraint and polymerizing at a given vertex, one can look for the solutions to the following set of equations (one per vertex)
\begin{equation}\label{convenient}
\left(-i\gamma\ell^2_{\rm Pl}\frac{\delta}{\delta  K_\varphi(x_j)
}\right) \Psi[M,K_{\varphi}(x_j)]= \pm  \frac{(x_j+2GM)}{
\sqrt{\frac{x_j}{x_j+2GM}+\frac{1}{4\gamma^2\rho^2}
	\sin^2\left(2\rho \gamma K_{\varphi}(x_j)\right)}}\Psi[M,K_{\varphi}(x_j)].
\end{equation}
The states are given by
\begin{equation}\label{109}
\Psi[M,K_{\varphi}(x_j)] =\exp\left(
\pm \frac{i}{\gamma\ell_{\rm Pl}^2}
f[K_{\varphi}(x_j),M]\right), 
\end{equation}
where
\begin{eqnarray}
f[M,K_{\varphi}(x_j)]&=& 
\frac{x_j+2GM}{4\gamma^2\rho^2}
F\left(\sin(2\gamma \rho K_{\varphi}(x_j)),
i\frac{x_j+2GM}{4\gamma^2\rho^2 x_j}\right),
\end{eqnarray}
is given in terms of $F(\phi,m)\equiv \int_0^\phi (1-m^2\sin^2 t)^{-1/2}dt$, the Jacobi
Elliptic function of the first kind.

Therefore, the complete solution is 
\begin{equation}\label{111}
\Psi[M,K_{\varphi}] =\prod_j^n\exp\left(
\pm \frac{i}{\ell_{\rm Pl}^2}
f[K_{\varphi}(x_j),M]\right). 
\end{equation}

We may notice that these states are normalizable with respect to the kinematical inner product. The observables of the model are simply the mass of the black hole $\hat M$ and its conjugate variable $\hat \tau = -i\hbar \partial_M$. The set of phase space functions $K_{\varphi}(x_j)$ for each vertex must be considered as a set of gauge parameters that tell us which Cauchy surface we are choosing (although other choices for parameterizing the Cauchy surfaces can also be considered). For instance, $K_{\varphi}(x_j)=0$ at any vertex $x_j$ yields the usual static Schwarzschild solution. Physical states are then given by
\begin{equation}\label{112}
|\Psi_{\rm phys}\rangle_{K_{\varphi}} =\int dM\,\Psi(M)\prod_j^n\exp\left(
\pm \frac{i}{\ell_{\rm Pl}^2}
f[K_{\varphi}(x_j),M]\right)|M\rangle, 
\end{equation}
and the physical inner product will be
\begin{equation}\label{113}
\langle \Psi_{\rm phys}|\tilde\Psi_{\rm phys}\rangle_{K_{\varphi}} =\int dM\,[\Psi_{\rm phys}(M)]^*\tilde \Psi_{\rm phys}(M). 
\end{equation}

It coincides with the inner product provided by Kuchar's quantization of Ref. \cite{kuchar}. 

For instance, the basic observables have a well-defined action on physical states
\begin{align}
\hat M  |\Psi_{\rm phys}\rangle_{K_{\varphi}} &=\int dM\,M\Psi(M)\prod_j^n\exp\left(
\pm \frac{i}{\ell_{\rm Pl}^2}
f[K_{\varphi}(x_j),M]\right)|M\rangle, \\
\hat \tau  |\Psi_{\rm phys}\rangle_{K_{\varphi}} &=\int dM\,(-i\hbar\partial_M)\left[\Psi(M)\prod_j^n\exp\left(
\pm \frac{i}{\ell_{\rm Pl}^2}
f[K_{\varphi}(x_j),M]\right)\right]|M\rangle.
\end{align}

We also have a well-defined set of operators corresponding to the triad on different vertices
\begin{align}\nonumber
\hat{E}^\varphi(x_j)|\Psi_{\rm phys}\rangle_{K_{\varphi}} =\pm  \int dM\,&\frac{(x_j+2GM)}{
	\sqrt{\frac{x_j}{x_j+2GM}+\frac{1}{4\gamma^2\rho^2}
		\sin^2\left(2\rho \gamma K_{\varphi}(x_j)\right)}}\Psi(M)\\
&\times\prod_j^n\exp\left(
\pm \frac{i}{\ell_{\rm Pl}^2}
f[K_{\varphi}(x_j),M]\right)|M\rangle.\label{eq:ephi-op}
\end{align}
Together with the gauge fixing, the other component of the triad $\hat E^x(x_j)= (x_j+2G\hat M)^2$ and its derivative $[\hat E^x(x_j)]'= 2(x_j+2G\hat M)$, we can construct any component of the metric and compute effective geometries for this quantum theory. 

It is remarkable that the resolution of the singularity has not been studied in this model. This will require a horizon penetrating slicing, like the one in Eddington---Finkelstein coordinates.  In addition, the loop quantization introduces a discretization in the spectrum of the kinematical operator $\hat{E}^\varphi(x_j)$ defined in Eq. \eqref{eq:ephi-kin} that should be present in the spectrum of the physical one given in Eq. \eqref{eq:ephi-op}.

\section{The Reissner-Nordstr\"om black hole}\label{sec:BH-MQ}

The quantization of spherically-symmetric spacetimes has been extended to charged black holes. Although they are expected not to be common in nature, they allow us to test several fundamental aspects of black hole physics, especially in the context of quantum gravity. They have already been considered before in Ref. \cite{mgp} in the context of spherical loop quantum gravity. 

Following the ideas of \cite{loukowintershilt,mgp}, we introduce a spherically-symmetric electromagnetic vector potential
${\bf A}= \Gamma dr+\Phi dt$ described by two configuration
variables $\Gamma,\Phi$ and the canonically-conjugate momenta,
$P_\Gamma,P_\Phi$, respectively. In order to avoid monopoles, we assume a trivial bundle for the electromagnetic field. In the Hamiltonian formalism, $\Phi$ plays the role of a Lagrange multiplier and will be ignored since it will be irrelevant in our description. The Hamiltonian turns out to be a linear combination of the Hamiltonian, diffeomorphism and the electromagnetic Gauss constraints. They are given by 
\begin{eqnarray}
H &=& \nonumber
G^{-1}\bigg(-\frac{E^\varphi}{2\sqrt{E^x}} - 2 K_\varphi \sqrt{E^x} K_x  
-\frac{E^\varphi K_\varphi^2}{2\sqrt{E^x}}+\frac{\left((E^x)'\right)^2}{8\sqrt{E^x}E^\varphi}
-\frac{\sqrt{E^x}(E^x)' (E^\varphi)'}{2 (E^\varphi)^2} +
\frac{\sqrt{E^x} (E^x)''}{2 E^\varphi} \\
&&+G \frac{E^\varphi}{2
	\left(E^x\right)^{3/2}} P_\Gamma^2\bigg),\\
C &=& G^{-1}\bigg(-(E^x)' K_x +E^\varphi (K_\varphi)'-G\, \Gamma P_\Gamma'\bigg),\\
{\cal G} &=& P_\Gamma',
\end{eqnarray}

In order to proceed with the quantization, in Ref. \cite{mgp} the authors adopt an Abelian Hamiltonian constraint. However, it is necessary to add appropriate boundary terms that can be inferred from the falloff conditions of the classical phase space functions. They were studied in Refs. \cite{saeed,kieferlouko}, and contribute with the well-known mass $M$ and, additionally, with the charge $Q$ of the electromagnetic field. In the following, they will be regarded as two global degrees of freedom. In addition, since we are interested in a static configuration, we will consider the gauge fixing $\Gamma=0$. The Gauss constraint (once the boundary terms have been introduced) yields $P_\Gamma=-Q$. 

With all of the previous considerations, the total Hamiltonian reads 
\begin{align}\nonumber
H_T =&G^{-1}\int dx \bigg\{ -N
\left(-\sqrt{E^x}\left(1+K_\varphi^2 +\frac{G Q^2}{E^x}\right)
+\frac{\left(\left(E^x\right)'\right)^2\sqrt{E^x}}
{4 \left(E^\varphi\right)^2}+2 G M
\right)'\\
&+ N_r \left[-
(E^x)' K_x +E^\varphi (K_\varphi)'\right]
\bigg\}. 
\end{align}

The contribution of the electromagnetic field is codified in the $Q^2$-term of the previous expression. Let us notice that the Reissner---Nordstr\"om metric in
Schwarzschild coordinates 
\begin{equation}
ds^2=-\left(1-\frac{2 G M}{x}+\frac{G Q^2}{x^2}\right)dt^2+
\frac{1}{1-\frac{2 G M}{x} + \frac{G Q^2}{x^2}}dr^2+x^2d\Omega^2,
\end{equation}
can be obtained form the previous canonical framework by means of the gauge fixing conditions $E^x=x^2$ and
$K_\varphi=0$. 

For the quantization of this theory, we will follow the prescriptions of Refs. \cite{gp-lett,mgp}. There, the basic elements of the quantum theory are given in terms of holonomies of the connection. We choose a kinematical Hilbert space given by
\begin{equation}
{\cal H}_{\rm kin}= {\cal H}^M_{\rm kin}\otimes {\cal H}^Q_{\rm kin} \left[\bigotimes_{j}^n \ell^2_j\otimes
\ell_j^2\right],
\end{equation}
such that $\ell^2_j$ is a set of Hilbert spaces of square summable functions, one associated with the edges and the other one with the vertices, of graphs of the form of Eq. \eqref{eq:kin-graph}.
Besides, ${\cal H}^M_{\rm kin}$ and ${\cal H}^Q_{\rm kin}$ are the Hilbert spaces associated with the mass
and charge of the spacetime. Here, we adopt a representation in terms of periodic functions in $K_\varphi(x)$ of period $\pi/\rho(x)$, with $\rho(x)$ a real-valued function. On this kinematical Hilbert space, the valences $\vec{\mu}$ are real numbers that take values on a semilattice of the form $\mu_j=\varepsilon_j+4\rho_j n_j$, where $\varepsilon_j$ is a phase associated to the periodicity conditions of the states $\psi(0)=e^{i\varepsilon_j}\psi(\pi/\rho_j)$. We could have also adopted a polymer representation on the Bohr compactification of the real line like the one discussed in Sec. \ref{sec:kinemat}. 

The inner product in this kinematical space is similar to the one in Eq. \eqref{eq:kin-inner-prod-qp}, but this time
\begin{align}\label{eq:kin-inner-prod-MQ}
\langle\vec{k},\vec{\mu},M,Q
|\vec{k}',\vec{\mu}',M',Q'
\rangle=\delta_{\vec{k},\vec{k}'}\delta_{\vec{n},\vec{n}'}\delta(M-M')\delta(Q-Q')\;.
\end{align}

The operator representation is similar to the one given in Eqs. \eqref{eq:M}, \eqref{eq:ex} and \eqref{eq:evarphi}, with the restrictions on the spectrum of $\hat E^\varphi$ mentioned before. In addition, the charge of the black hole in this representation is given by
\begin{equation}
{\hat{Q} } |\vec{k},\vec{\mu},M,Q\rangle
= Q |\vec{k},\vec{\mu},M,Q\rangle.\label{eq:Q}
\end{equation}

In order to continue with the quantization of the system, we will come back to the classical scalar constraint, that will be written as
\begin{equation}
H(N)= \int dx N' H_+ H_-,
\end{equation}
with 
\begin{equation}
H_{\pm} =\sqrt{\sqrt{E^x}\left(1 +K_\varphi^2
	+\frac{G Q^2}{E^x}\right)-2G M }
\pm \frac{\left(E^x\right)'\left(E^x\right)^{1/4}}{2 E^\varphi}.
\end{equation}
We then now redefine the lapse function by absorbing one of the two factors together with a final rescaling by a factor of $4 \left(E^\varphi\right)^2$. The final form of the scalar constraint will be
\begin{equation}
H(\bar{N})=\int dx \bar{N}\left(2 E^\varphi
\sqrt{\sqrt{E^x}\left(1+K_\varphi^2+\frac{GQ^2}{E^x}\right)-2 G M}-{\left(E^x\right)'\left(E^x\right)^{1/4}}\right).  
\end{equation}

This operator admits a simple representation in the space of cylindrical functions of the connection $K_\varphi$ given by
\begin{equation}
\hat{H}(\bar{N}) \vert \Psi\rangle =
\int dx \bar{N}\left(2 
\left[\sqrt{\widehat{\sqrt{|\hat E^x|}}\left(1+\frac{\widehat{\sin^2\left(2\rho
			\gamma\hat K_\varphi\right)}}{4\gamma^2\rho^2}+\frac{G\hat Q^2}{\hat{E^x}}\right)-2 G
	\hat M}\right]
\hat{E}^\varphi
-{\widehat{\left(\hat E^x\right)'}\widehat{\left|\hat E^x\right|^{1/4}}}\right)  \vert \Psi\rangle, 
\end{equation}
where this factor ordering has been chosen for the sake of simplicity. The solutions to the constraint can be computed explicitly. They admit a factorization on each vertex, since the constraint has been selected in such a way that it does not create new vertices nor edges. At each vertex, they fulfill the equation
\begin{equation}
2 i \gamma\ell_{\rm Pl}^2 \frac{\sqrt{1+m_j^2 \sin^2 \left(y_j\right)}}{m_j}
\partial_{y_j} \psi_j 
+\gamma\ell_{\rm Pl}^2 \left(k_j-k_{j-1}\right)\psi_j=0.
\end{equation}
where $y_j=2\rho \gamma K_\varphi(x_j)$ and 
\begin{equation}
m_j^2=(2\gamma\rho)^{-2}\left(1-\frac{2GM}{\sqrt{\ell_{\rm Pl}^2
		k_j}}+\frac{G Q^2}{\ell_{\rm Pl}^2k_j}\right).
\end{equation}

The solutions to these set of equations are give by
\begin{equation}
\psi_j\left(M,Q,k_j,k_{j-1},K_\varphi\left(x_j\right)\right)=
\exp\left(\frac{i}{2} m_j \left(k_j-k_{j-1}\right)F\left(2\rho\gamma
K_\varphi\left(x_j\right),i m_j\right)\right),
\end{equation}
where $F$ is defined as,
\begin{equation}
F\left(\phi,K\right)=\int_0^\phi \frac{dt}{\sqrt{1+K^2 \sin^2 t}}.
\end{equation}
Here, $m_j$ is chosen to be a purely imaginary number in the upper complex plane if $m_j^2<0$ (a vertex inside the black hole between the two horizons). One can see that these states are normalizable with respect to the kinematical inner
product.

The last step of the quantization is to identify the solutions to the scalar constraint that are also invariant under the group of spatial diffeomorphisms generated by the classical diffeomorphism constraint. For a brief discussion, see Sec. \ref{sec:physical}. 

The resulting states are endowed with a well-defined inner product. This allows us to identify quantum Dirac observables. The mass $\hat M$ and the charge $\hat Q$ are two of them, with classical Dirac analogues. The third one  is not related to any classical Dirac observable. It corresponds to the same emerging Dirac observable defined already in Eq. \eqref{eq:ObsO}. Together with suitable parameters, we can construct a full canonical quantization. For instance, as we did in Sec. \ref{sec:physical}, a possible choice is to identify again here $z\to z(x)$, such that it codifies the dependence of phase space functions with respect to the spatial diffeomorphisms, and $K_\varphi\to K_\varphi(x)$ with another functional parameter, labeling different spatial slices.  Therefore, we can promote phase space functions to parametrized Dirac observables. For instance, the triad components take the form
\begin{align}
\hat{E}^x(x) \vert \Psi\rangle_{\rm phys} &=\hat{O}\left(z(x)\right) \vert \Psi\rangle_{\rm phys},\\
\hat{E}^\varphi(x) \vert \Psi\rangle_{\rm phys} &=\frac{{\widehat{\left(\hat E^x\right)'}\widehat{\left|\hat E^x\right|^{1/4}}}}{2 
	\left[\sqrt{\widehat{\sqrt{|\hat E^x|}}\left(1+\frac{{\sin^2\left(2\rho\gamma
				\hat K_\varphi\right)}}{4\gamma^2\rho^2}+\frac{G\hat Q^2}{\hat{E^x}}\right)-2 G
		\hat M}\right]}\vert \Psi\rangle_{\rm phys}.
\end{align}
Out of them, we can construct the operators corresponding to the metric components as parametrized Dirac observables as
\begin{eqnarray}
\hat g_{tx} &=& - \frac{\frac{{\sin\left(2\rho \gamma
			\hat K_\varphi\right)}}{2\gamma\rho} \left(\hat E^x\right)'}
{2 \sqrt{|\hat E^x|} \sqrt{\left(1+\frac{{\sin^2\left(2\rho\gamma
				\hat K_\varphi\right)}}{4\gamma^2\rho^2}\right) -\frac{2 G \hat M}{\sqrt{|\hat E^x|}} +
		\frac{G \hat Q^2}{\hat E^x}}},\\
\hat g_{xx} &=& 
\frac{\left(\left(\hat E^x\right)'\right)^2}
{4 \hat E^x \left(\left(1+\frac{{\sin^2\left(2\rho\gamma
			\hat K_\varphi\right)}}{4\gamma^2\rho^2}\right) -\frac{2 G \hat M}{\sqrt{|\hat E^x|}} +
	\frac{G \hat Q^2}{\hat E^x}\right)},\\
\hat g_{tt} &=& -\left(1-\frac{2 G \hat M}{\sqrt{|\hat E^x|}}+\frac{G \hat Q^2}{\hat E^x}\right).
\end{eqnarray}
Let us notice that in the quantum theory, the lapse and shift are treated as classical functions. Therefore, it is not clear how to specify them or to promote to quantum operators. In this case, we simply take classical expressions and directly promote the corresponding phase space functions to parametrized Dirac observables. This procedure will yield to factor ordering and polymer schemes ambiguities. This is why the explicit expression we give here can differ from the operators considered in Ref. \cite{mgp}.

This quantization is singularity free. However, following Ref. \cite{mgp}, the self-adjointness of the metric components cannot be used here to rule out vertices with $k_j=0$ from the physical Hilbert space. The only argument that has been considered by now is that the constraint is diagonal with respect to the observable $\hat O(z)$, which allows one to restrict the study to states with $k_j\neq 0$. 

This quantum description may have important implications with respect to the stability of the Cauchy horizons of the
interior of Reissner---Nordstr\"om black holes \cite{instability}. In classical general relativity, the instability of the horizons is due to the following heuristic argument. Let us consider an observer that enters the black hole while another observer is in the exterior. By the time the first observer reaches the interior, the second one reaches $i^+$. Therefore, the observer in the exterior can send a huge amount of energy that is received by the former observer in a finite time interval. At the quantum level, the spectrum of $\hat{E}^x$ is discrete. Then, there is a minimum spacing in the variable $\hat R = \sqrt{|\hat E^x|}$ when we jump from one vertex to its neighbor. This minimum is equal to $\ell_{\rm Pl}^2/(2 \hat R)$. For instance, the spacing of the geometry close to the horizon of a black hole is of the order of $\ell_{\rm Pl}^2/(4 G M)$. Therefore, trans-Planckian modes of arbitrarily high frequency are eliminated. In addition, the modes that propagate in the exterior of the black hole cannot be blue-shifted indefinitely close to the horizon. They are indeed attenuated and reflected. Therefore, not all the energy crosses the horizon and reaches the infalling observer. Some of it is backscattered towards $scri^+$, while it will be visible only for remote future external observers. All of these arguments indicate that the physical picture of the classical theory changes within the quantum description. However, a detailed calculation will be needed in order to understand in depth the concrete improvements that it provides. 

\section{Quantum test fields on quantum geometries}\label{test-f}

Quantum field theories on these effective quantum spacetimes have not been studied in depth. If one considers quantum test linear field theories on classical geometries, it is well known that nonlinear observables, like the stress-energy tensor, require the adoption of regularization and renormalization schemes (due essentially to some of the divergences present in these type of quantizations). However, it is expected that in a genuine quantum theory of gravity, where geometry, fields and their interactions are described at the quantum level, no divergences will appear. Indeed, in Ref. \cite{gop-casimir}, a quantum field theory on these quantum spacetimes was studied, showing that it is not necessary to renormalize in order to get the right answer. In addition, Ref. \cite{gp-semi} provides a study about Hawking radiation and the corrections with respect to the traditional result due to quantum geometry phenomena. One of the issues that deserves special attention is the covariance of these scenarios. Indeed, the fundamental discreteness together with the test field approximation yields a combination that is not covariant at Planck scales. However, those Planckian modes correspond to energies where the test field approximation is lost, and therefore, the physics they involve cannot be fully trusted.

\subsection{Casimir effect}

Let us focus now in Ref. \cite{gop-casimir}, where the authors consider the Casimir effect of a scalar field between two concentric spherical static shells where the field vanishes. In this case, one considers a spin network with radial positions at $r^2_i=\ell_{\rm Pl}^2 k_i$, where the difference between two successive values of the radial coordinate $r_j$ is at least $\ell_{\rm Pl}^2/(2r_j)$. We choose a concrete spin network such that $r_j=r_0+j \Delta$ in the interior of the slab between two spherical shells of radii $r_0$ and $r_0+L$, with $j$ a suitable integer labeling the vertices and $\Delta$ their separation. The number of vertices on each region is given by the quotient of its length over the step. Inside the slab of length $L$ we have $N_L=L/\Delta$ vertices. The scalar field vanishes at the shells as is usual
for calculations of the Casimir effect. Besides, the authors assume that their radii are in the asymptotic region where the background quantum metric is flat, i.e., $r_0\gg 2M$. For simplicity, all the calculations are carried out with
respect to an observer at rest in $r=r_0$. In addition, contributions of the field outside the shells must be also considered. In this sense one introduces two auxiliary shells at $r_0+L_0$ and $r_0-L_1$, such that $L_0\gg L$. The value of
$L_1$ can be selected arbitrarily, covering a large portion of the space-time, but still in the asymptotic region. 

Then, one can adopt an expansion in Fourier modes for the scalar field 
$u_{n,\ell,m} = \exp(-i\omega
t)R_{\ell}(\omega,r_j)Y_{\ell,m}(\theta,\varphi)/(\sqrt{2\pi\omega}r_j)$, and solve its equations of motion.  The time-dependent and the angular parts of the equations of motion are well known, but the radial modes have to be determined. They are not known in closed form; but, in Ref. \cite{gop-casimir}, they were studied numerically, and approximated expressions are available. 

For instance, an approximated dispersion relation was provided
\begin{equation}\label{eq:disp-rel}
\omega_{{\frak n},\ell}^2\simeq\frac{4}{\Delta^2}\sin^2\left(\frac{\Delta
	k_{\frak n}}{2}\right)+\frac{\ell(\ell+1)}{r_0^2},
\end{equation}
with $k_{\frak n}=({\frak n}\pi)/(N_L\Delta)$ the wave number of each mode. The mode number ${\frak n}$ is bounded above due to the discretization of the radial coordinate inside the slab between the shells, while $\ell$ is bounded above since the dominant contributions are due to modes of angular momentum $\ell$ lower than $2r_0/\Delta$ (the maximum mode frequency on the lattice times the radius $r_0$). This bound is due to the fact that the solutions are considerably suppressed in the interior of the slab due to the centrifugal potential and are disregarded in the calculations.

The stress-energy tensor components can be easily computed once the Green's
function associated with the slab $G_+^L(x,x')=\langle 0_L\vert \phi(x)\phi(x')\vert
0_L\rangle$ is known. After some calculations, one can see that
\begin{equation}
G_+^L\left(x;x'\right)\simeq\frac{1}{L}\sum_{{\frak n}=1}^{N_L-1}\sum_{\ell=0}^{\frac{2r_0}{\Delta}}\sum_{m=-\ell}^\ell
\frac{ e^{-i\omega_{{\frak n},\ell} \left(t-t'\right)}}{\omega_{{\frak n},\ell}}\frac{\sin\left(k_{\frak n}z\right)}{r_j}\frac{\sin\left(k_{\frak n}z'\right)}{r_{j'}}Y_{\ell,m}\left(\theta,\varphi\right)
Y^*_{\ell,m}\left(\theta',\varphi'\right),
\end{equation}
with $z=r-r_0$, and similarly for $z'$, and $\omega_{{\frak n},\ell}$ is given in (\ref{eq:disp-rel}).

Then, in order to compute the Casimir force between the two shells, it is only necessary to compute the $T_{00}$-component of the stress-energy tensor, which is basically given by 
\begin{align}
\langle 0_L\vert T_{00}\vert 0_L\rangle =	\left.  \frac{\partial^2}{\partial t\, \partial t'}
G_+^L\right\vert_{x=x'},
\end{align}
where $\vert 0\rangle_L$ is the vacuum state for the slab of width $L$. 

For the scalar field inside the other regions, the same construction can be adopted without additional considerations, and the corresponding expectation values of the $T_{00}$-component of the stress-energy tensor $\langle 0_{\tilde L_0}\vert T_{00}\vert 0_{\tilde L_0}\rangle$, with the separation $\tilde L_0=(L_0-L)$, and $\langle 0_{L_1}\vert T_{00}\vert 0_{L_1}\rangle$ can be incorporated in the calculation. 

The Casimir force is given as (minus) the derivative of the energy of the system with respect to $L$. In the limit 
$L_0\gg L$ and for small $\Delta$, one gets
\begin{eqnarray}\nonumber
{\rm Force}&=& -\frac{d}{dL}\left(\int^0_{-L_1} dz \langle 0_{L_1}\vert T_{00}\vert
0_{L_1}\rangle+\int_0^L dz \langle 0_L\vert T_{00}\vert
0_L\rangle +\int_L^{L_0} dz \langle 0_{\tilde L_0}\vert T_{00}\vert 0_{\tilde
	L_0}\rangle\right)\\
&=&-\frac{\pi^2}{480L^4}+{\cal
	O}(L_0^{-1})+{\cal
	O}(\Delta).
\end{eqnarray}

It coincides with the exact result in the planar case \cite{ozcan,fulling} at leading order, since the calculations are done at the asymptotic region with $L\ll r_0$. Besides, the improved stress-energy tensor proposed in Refs. \cite{callancolemanjackiw,fulling} agrees with this result. An important remark about this calculation is the fact that it is obtained as the variation of the energy of a closed system, without any subtraction of an hypothetical external vacuum energy. The energy of both slabs are indeed added, rather than subtracted, in order to achieve a meaningful physical result. In addition, let us emphasize that it is not necessary to adopt any regularization and/or renormalization scheme, as in the continuum limit.

\subsection{Hawking radiation}

Another interesting phenomenon intimately related with black hole physics is the Hawking radiation process. It has been studied in Ref. \cite{gp-semi}. In this case, the authors consider a massless scalar field propagating in an effective quantum geometry within the loop quantization presented in this review. As we saw above, these effective geometries naturally regularize the matter content by suppressing trans-Planckian modes. 

Although we are not going to show here the details (we encourage the readers to check Ref. \cite{gp-semi}), the Boulware, Hartle---Hawking and Unruh vacuum states can be computed on this quantum geometry (it is only necessary to formulate the corresponding coordinates on it). We will mainly focus here on the computation of the number of particles of the state at the future scri with respect to the one at the past horizon.

In this situation, one adopts the typical calculations to the case in which there is a Planck scale cutoff (see, for instance, Ref. \cite{salas-agullo}). We start with the two sets of modes
\begin{eqnarray}
u^{\rm in}_{\omega,\ell,m} &=&\frac{1}{\sqrt{4 \pi
		\omega}}\frac{\exp\left(-i\omega U\right)}{r} Y_\ell^m\left(\theta,\varphi\right),\\
u^{\rm out}_{\omega,\ell,m} &=&\frac{1}{\sqrt{4 \pi
		\omega}}\frac{\exp\left(-i\omega u\right)}{r} Y_\ell^m\left(\theta,\varphi\right),
\end{eqnarray}
where $u$ is the usual null coordinate that depends on the time of an observer in the spatial asymptotic region and the tortoise radial coordinate, while $U$ is the standard null coordinate in the Kruskal extension of a black spacetime. We will introduce the notation ${\bf n}=\{\omega,\ell,m\}$ for the sake of simplicity. The expectation value of the ``out'' number operator on the ``in'' vacuum can be expressed in terms of these mode functions
\begin{equation}
\langle {\rm in}\vert N^{\rm out}_{{\bf n}_1,{\bf n}_2}\vert {\rm in}\rangle =-\sum_{\bf n} \left(u_{{\bf n}_1}^{\rm out},{u_{\bf n}^{\rm in}}^*\right)\left({u_{{\bf n}_2}^{\rm out}}^*,{u_{\bf n}^{\rm in}}\right).
\end{equation}

In order to compute the previous summation, it is necessary to introduce an exponential cutoff in $\omega$ of the form $e^{-\epsilon\omega}$. The parameter $\epsilon$ cannot be taken to vanish at the end of the calculation and in Ref. \cite{gp-semi} is chosen to be $\epsilon\simeq \ell_{\rm Pl}$. It allows us to carry our the calculation explicitly, as in the continuum theory, since the dispersion relation in the lattice is very well approximated for the relevant frequencies by the continuum one. 

With these considerations in mind, and carrying out the summation in ${\bf n}$, the number operator takes the form
\begin{equation}
\langle {\rm in}\vert N^{\rm out}_{{\bf n}_1,{\bf n}_2}\vert {\rm in}\rangle = A
\int_{I+} dU_1 dU_2 \frac{\exp\left(-i\omega_1 u_1(U_1)+i\omega_2
	u_2(U_2)\right)}
{\left(U_1-U_2-i\epsilon\right)^2},
\end{equation}
with
\begin{equation}
A=\frac{t_{\ell_1}\left(\omega_1\right)t^*_{\ell_2}\left(\omega_2\right)}
{4\pi^2 \sqrt{\omega_1\omega_2}}\delta_{\ell_1,\ell_2}\delta_{m_1,m_2},
\end{equation}
where the factors
$t_\ell(\omega)$ are the ``transmission coefficients'' that appear due to the potential in the radial wave equation. 

Under the change of variables
\begin{eqnarray}
U_1&=&-4 G M \exp\left(\frac{u_M+z}{4 G M}\right),\\
U_2&=&-4 G M \exp\left(\frac{u_M+z}{4 G M}\right),
\end{eqnarray}
with $u_M=u_1+u_2$ and $z=u_2-u_1$, the previous integral reduces to one in $z$ that can be computed explicitly. In the continuum theory, one gets the usual Hawking formula
\begin{equation}
\langle {\rm in}\vert N^{\rm out}_{{\bf n}_1,{\bf n}_2}\vert {\rm in}\rangle =
\frac{\vert t_\ell(\omega_1)\vert^2}{\exp\left(2 \pi \omega_1/\kappa\right)-1}\delta({\bf n}_1,{\bf n}_2).
\end{equation}
where $\kappa=1/(4 G M)$. However, within our quantum background, there is a cutoff of the order of $\ell_{\rm Pl}$ in the variable $u$. One therefore has
\begin{equation}
\left(u_1-u_2\right)^2 \ge {\ell_{\rm Pl}^2},
\end{equation}
which is the same as that of Ref. \cite{salas-agullo}. 
This leads to the same formula found there
\begin{equation}
\langle {\rm in}\vert N^{\rm out}_{{\bf n}_1,{\bf n}_2}\vert {\rm in}\rangle =\left(
\frac{\vert t_\ell(\omega_1)\vert^2}{\exp\left(2 \pi \omega_1/\kappa\right)-1}
-\frac{\kappa^2 \ell_{\rm Pl}}{96 \pi^3 \omega_1}\right)\delta({\bf n}_1,{\bf n}_2).
\end{equation}
We see that the cutoff induced by the discrete geometry introduces a correction much smaller than the leading contribution, at least for not very high frequencies and black holes with a Schwarzschild radius bigger than the Planck length. 

\section{Coupling to a spherical null dust shell}\label{sec:shell}

One of the most interesting generalization of the previous results is to extend the quantization when the geometry is coupled to matter. It would be possible to study gravitational collapse within the quantum realm. On the one hand, the coupling of matter in loop quantum gravity is not well understood yet (see Refs. \cite{lqg,lqg1,lqg2}). On the other hand, spherically-symmetric gravitational collapse shows a rich dynamics even at the classical level, like black hole formation and critical phenomena \cite{choptuik}. Several effective descriptions \cite{lqg-coll,lqg-coll1,lqg-coll2,lqg-coll3} incorporate, in a well motivated, but heuristical sense, loop quantum gravity corrections; usually the ones from the inverse of the triad operator and/or those coming from the effective Friedmann equation of isotropic loop quantum cosmology \cite{lqc,lqc1,lqc2}. These effective scenarios agree with respect to the resolution of the classical singularity.

One of the simplest models where dynamical black hole formation can be studied corresponds to a matter content consisting of a spherical null dust thin shell. The dynamics of a test thin null dust shell on a quantum geometry was studied in Ref. \cite{gp-shell}. There, the shells propagate through a high curvature region where the singularity used to be and emerges in a different asymptotic region corresponding to the pre-existing white hole quantum space-time. The quantum packets of the shells are distorted when they cross the high quantum region that replaces the singularity. It is remarkable that the evolution of the shells is entirely unitary. The self-gravitating shell (with no test field approximation) has been studied by many authors \cite{kuchar-shell,louko,hajicek,haji-kiefer, corichi}. However, we will follow here the main ideas about the classical model in Ref. \cite{louko}. It is worth commenting that it has been partially studied at the quantum level in Refs. \cite{hajicek,haji-kiefer, corichi}. In these cases, a classical reduction yields a true Hamiltonian ruling the dynamics of the shell. The resulting model (only covers the region close to the shell) is simple enough that a standard quantization can be adopted. For instance, the quantization carried out in Ref. \cite{corichi} shows that quantum corrections modify the dynamics in the high curvature region, such that the shell, once it approaches this regime, bounces and expands into a different asymptotic region. On the other hand, the quantization of Refs. \cite{hajicek,haji-kiefer}, under the requirement of self-adjoints of the true Hamiltonian (i.e., a natural boundary condition), yields a quantum dynamics where the shell bounces very fast when it approaches its Schwarzschild radius, even for macroscopic black holes. This is due to the fact that the eigenstates of the quantum true Hamiltonian are a superposition of both black and white hole horizons. It is the so-called grey horizon \cite{hajicek}.

In the quantization presented in Ref. \cite{cgop}, it was shown that for this particular model, it is possible to prove that the Hamiltonian constraint, at the classical level, can be Abelianized as in the vacuum model. Besides, despite the dynamics is nontrivial, it is possible to identify Dirac observables in the model; concretely, the mass of the shell $m$ and its conjugate momentum $V$, which turns out to be the time in Eddington---Finkelstein coordinates. Within these coordinates, the spacetime metric can be explicitly constructed, in terms of two parameters: the radial coordinate $x$ and the time of an asymptotic observer $t$, as well as the two Dirac observables $m$ and $V$.

One can then proceed with the quantization of the model. For instance, one can choose a polymer representation like the one adopted in Sec. \ref{sec:BH-MQ}. The scalar constraint is represented by an operator that preserves the number of vertices and edges. However, more important, and one of the nontrivial technical results of Ref. \cite{cgop}, is the fact that a suitable choice of factor ordering, polymerization and discretization allow one to construct a quantum constraint free of anomalies. It is worth commenting, however, that the polymerization that is compatible with an abelian Hamiltonian constraint in presence of matter involves corrections that were unnecessary in the vacuum case, and therefore, they were ignored. Indeed, these polymer corrections can be naively identified with a classical canonical transformation of the form $(K_\varphi,E^\varphi)\to\left(\sin\left(\rho K_\varphi\right)/\rho, E^\varphi/\cos\left(\rho K_\varphi\right)\right)$. As we see, it is not well defined when the cosine vanishes. The Hamiltonian constraint, in the present model, is a well-defined operator almost everywhere in the kinematical Hilbert space; concretely, in those subspaces with support on  $\mu_j=2\rho_j(l_j+\delta_j)$ with $l_j$ an integer and $\delta_j$ in $(0,1)$. The Hamiltonian constraint, as a kinematical operator, does not mix states with support on different $\delta_j$. If one restricts the study to any of these subspaces, it is likely that one can find a consistent (singularity free) quantization of the model. However, the fact that the constraint is not well defined for $\delta_j=0,1$ also indicates the pathologies of the holonomy corrections involved in the construction, since they can have consequences regarding the singularity resolution in those sectors. However, only an exhaustive study of the model can shed light on all of the previous points.  One must keep in mind that this quantization opens the possibility for the study of full spherically-symmetric quantum collapse in loop quantum gravity. For instance, it is possible to find at the algebraic level quantum operators that commute with the constraints. However, it is has not been possible to find a realization for them as self-adjoint operators in the polymer space. Neither the solutions to the constraints have been studied yet carefully. 

However, the strategy suggested in Ref. \cite{cgop} is to assume a standard self-adjoint representation of the observables. One will lose some of the corrections coming from loop quantum gravity, but keeping some of the important ones. For instance, for diffeomorphism-invariant states the observable $\hat O(z)$ of the vacuum models is also present here. Then, the geometry is quantized. Indeed, the parametrized observables in Eddington---Finkelstein coordinates could be represented in the physical space. Although in Ref. \cite{cgop} they are provided for a particular choice of coordinates, it is not difficult to carry out the calculation for any other choices admissible in the canonical theory. In this sense, the covariance is not violated. The final physical quantum picture corresponds to a discretization of the classical theory. The geometries can be extremely smooth if one considers superpositions of spin networks highly peaked on a quantum geometry, as well as superpositions of the mass. In this situation, the high curvature region, including the classical singularity, is replaced by a portion of the spacetime that is not well approximated by smooth geometries, but that is still regular. In this sense, following the ideas of Secs. \ref{sec:physical} and \ref{sec:BH-MQ}, the singularity is eliminated. This picture is in agreement with Ref. \cite{corichi}. The shell, if the original flat topology possesses two asymptotic regions, can transverse this high curvature regime and emerge in the other asymptotic region, leaving behind a white hole. If the initial topology has only one asymptotic region, the shell will reach the deep quantum regime. However, it is not known yet if the shell will bounce back or if it will be trapped in the interior indefinitely. If the shell bounces back, global quantum gravity effects were postulated to be able to propagate at low curvature regimes \cite{rovelli,rovelli1,rovelli2,bcg,bcg1,bcgj}. If this is the case, it is possible that some imprints could be observable \cite{rovelli1,rovelli2}. However, there is no agreement at present in the typical times of black hole evaporation, whether they are long \cite{rov-hagg} or not \cite{bcg,bcg1,bcgj}.

\section{Conclusions}\label{sec:concl}

In summary, in this contribution, we have reviewed the recent advances in the loop quantization of spherically-symmetric spacetimes and their application in several aspects of black hole physics like singularity resolution, black hole formation and interesting phenomena on quantum field theory on quantum spacetimes. Concretely, we reviewed the main aspects that are usually considered in order to carry out the full quantization of a vacuum spherically-symmetric spacetime following the Dirac quantization approach. We start with a description of classical general relativity in terms of real Ashtekar---Barbero variables \cite{lqg}. We then impose spherical symmetry as it was originally done in Ref. \cite{bengtsson}. The resulting 1+1 theory has a dynamics ruled by a Hamiltonian constrained to vanish. It is a linear combination of a momentum constraint that generates spatial diffeomorphisms and a scalar constraint that also generates time reparametrizations. In vacuum, there is only one degree of freedom: the ADM mass \cite{kuchar}. In presence of matter, the dynamics is highly nontrivial. In both cases, the constraint algebra shows structure functions. This is an additional handicap if one is tempted to adopt a canonical Dirac quantization. However, since loop quantum gravity is based on it, we have to face this issue somehow. The way it is avoided was suggested in Ref. \cite{gp-lett}, and it consists of a redefinition of the scalar constraint and the lapse and shift functions in such a way that the new scalar constraint is a total derivative. This trick seems to work in the presence of different types of matter \cite{bbr,cgop} and other midi-superspace models in vacuum \cite{bb,gowdy-LRS}. At the classical level, this transformation is in agreement with the solution space of the original set of constraints. The only potential problems appear if one imposes homogeneity. However, in this case, the original set of constraints gives a successful description, and no Abelianization is required. 

Then, we show how a loop representation for the vacuum model allows us to construct a complete quantization via group averaging techniques based on the Dirac quantization. Then, the quantum dynamics can be analyzed by means of relational observables \cite{gop,gop2}. It is singularity free, and smooth geometries can be constructed such that they cover very large portions of the corresponding classical continuous spacetimes. At the high curvature regime, the quantum fluctuations do not provide smooth geometries, but the corresponding regions are regular. We compare this quantization with previous approaches \cite{ash-bojo,cgp,cgp2,cs} where a classical reduction is adopted before quantizing, in order to check what are the common physical predictions. In all of these cases, one recovers smooth semiclassical geometries at low curvatures and a regular description without divergences at the high curvature region. We also describe how this quantization can be extended to Reissner---Nordstr\"om black holes \cite{mgp}. Here we obtain the same conclusions regarding the semiclassical smooth geometries and the resolution of the singularity. 

On the resulting effective geometries, it has been possible to study phenomena of quantum field theories on quantum spacetimes \cite{gp-semi,gop-casimir}. In these cases, the main effective correction coming from quantum geometry is the natural regularization that the discretization induces on quantum field theories. Let us comment that such discretization violates the covariance of the field theory at Planck scales. However, one has to take into account that in physically-relevant situations the discretization can be chosen considerable smaller than the Planck length, so that the covariance is preserved for all practical purposes. Nevertheless, it is important to realize that the covariance is broken since the system is not fully solved in the context of quantum gravity. In other words, we believe that the covariance is broken since one is working in the test field approximation. Nevertheless, one can analyze several physical phenomena, like the Casimir effect \cite{gop-casimir} or Hawking radiation \cite{gp-semi} of quantum test fields on quantum geometries. One of the common characteristics is the fact that the discretization introduces a regularization that allows one to work with finite expectation values of the stress-energy tensor, although in some situations, some kind of renormalization is still necessary since these finite values are of Planck order. Fortunately, in the case of the Casimir force, a natural subtraction in the computation yields the correct result without requiring additional renormalization. Regarding the Hawking radiation \cite{gp-semi}, for those frequencies that are not of the order of the Planck length and macroscopic black holes, one recovers the usual leading order contribution with an additional correction that is more relevant for smaller black holes. 

Finally, we comment on the most recent advances about spherically-symmetric spacetimes coupled to matter. In Ref. \cite{cgop} it was possible to construct a quantum scalar constraint with an algebra free of anomalies. The quantization gives effective geometries that are time-dependent and where, again, smooth and regular spacetimes emerge in some sectors of the physical Hilbert space. Nevertheless, the quantization has not been fully completed in the polymer representation, and further research is required in order to fully understand the quantum dynamics of this midi-superspace model. Since one can expect that a standard representation for those observables will approximate very well some of the true quantum states in the full polymer representation, it is possible to combine both of them, completing the quantization. The physical states are such that the effective spacetimes approximate very well smooth geometries, but they present a fundamental discretization. In addition, the deep curvature regime is free fo singularities where the corresponding portion of the effective spacetime, although it is regular, it cannot be described by a smooth geometry. Further comparisons must be done with the model of Refs. \cite{kiefer,haji-kiefer} in order to contrast the deep quantum dynamics of both models and see if they agree.

In summary, there have been promising advances in the study of midi-superspace models of gravity, but still much work has to be carried out in order to increase our understanding about the physical and technical aspects behind them, as well as their consequences and relation to the full theory.

\acknowledgments{The author acknowledges Jorge Pullin and the two anonymous Referees for their suggestions. This work has been supported by the grants FIS2014-54800-C2-2-P (Spain), NSF-PHY-1305000 (USA) and Pedeciba (Uruguay).}

\renewcommand\bibname{References}

\end{document}